\documentclass[fleqn,usenatbib]{mnras}

\usepackage[T1]{fontenc}
\usepackage{ae,aecompl}

\usepackage{times} % for details on the use of the package, please
\usepackage{tikz}
\usetikzlibrary{positioning}
\usepackage{xspace,ulem}

\usetikzlibrary{backgrounds}
\usetikzlibrary{arrows}
\usetikzlibrary{calc}
\usepackage{booktabs}
\usepackage{graphicx}	% Including figure files
\usepackage{amsmath}	% Advanced maths commands
\usepackage{amssymb}	% Extra maths symbols
\usepackage{times}
\usepackage{newtxtext}
\usepackage[varvw]{newtxmath}

\newcommand{\Cesar}[1]{\textcolor{green!55!blue}{#1}}

\title[Galaxy clusters in DGP gravity]{A general framework to test gravity using galaxy clusters IV: \\ Cluster and halo properties in DGP gravity
}

\author[M. A. Mitchell et al.]{
Myles A. Mitchell,$^{1}$\thanks{E-mail: m.a.mitchell@durham.ac.uk}
C\'esar Hern\'andez-Aguayo,$^{2,3,1}$
Christian Arnold$^{1}$
and Baojiu Li$^{1}$
\\
$^{1}$Institute for Computational Cosmology, Department of Physics, Durham University, South Road, Durham DH1 3LE, UK\\
$^{2}$Max-Planck-Institut f\"ur Astrophysik, Karl-Schwarzschild-Str. 1, D-85748, Garching, Germany\\
$^{3}$Excellence Cluster ORIGINS, Boltzmannstrasse 2, D-85748 Garching, Germany\\
}

\date{Accepted XXX. Received YYY; in original form ZZZ}

\pubyear{2021}

\begin{document}
\label{firstpage}
\pagerange{\pageref{firstpage}--\pageref{lastpage}}
\maketitle

% Abstract of the paper
\begin{abstract}
We study and model the properties of galaxy clusters in the normal-branch Dvali-Gabadadze-Porrati (nDGP) model of gravity, which is representative of a wide class of theories which exhibit the Vainshtein screening mechanism. Using the first cosmological simulations which incorporate both full baryonic physics and nDGP, we find that, despite being efficiently screened within clusters, the fifth force can raise the temperature of the intra-cluster gas, affecting the scaling relations between the cluster mass and three observable mass proxies: the gas temperature, the Compton $Y$-parameter of the Sunyaev-Zel'dovich effect and the X-ray analogue of the $Y$-parameter. Therefore, unless properly accounted for, this could lead to biased measurements of the cluster mass in tests that make use of cluster observations, such as cluster number counts, to probe gravity. Using a suite of dark-matter-only simulations, which span a wide range of box sizes and resolutions, and which feature very different strengths of the fifth force, we also calibrate general fitting formulae which can reproduce the nDGP halo concentration at percent accuracy for $0\leq z\leq1$, and halo mass function with $\lesssim3\%$ accuracy at $0\leq z\leq1$ (increasing to $\lesssim5\%$ for $1\leq z\leq 2$), over a halo mass range spanning four orders of magnitude. Our model for the concentration can be used for converting between halo mass overdensities and predicting statistics such as the nonlinear matter power spectrum. The results of this work will form part of a framework for unbiased constraints of gravity using the data from ongoing and upcoming cluster surveys.
\end{abstract}

% Select between one and six entries from the list of approved keywords.
% Don't make up new ones.

%%%%%%%%%%%%%%%%%%%%%%%%%%%%%%%%%%%%%%%%%%%%%%%%%%

%%%%%%%%%%%%%%%%% BODY OF PAPER %%%%%%%%%%%%%%%%%%

\section{Introduction}
\label{sec:introduction}

Galaxy clusters are the largest known gravitationally-bound objects in the Universe, and consequently are believed to have formed from the highest peaks of the primordial density perturbations. The global properties of clusters, such as their abundance, are thus powerful probes of cosmological models that influence the growth %and formation 
of cosmic structure. In particular, clusters can be used to study the behaviour of gravity on large scales, which is a key step towards explaining phenomena such as the late-time accelerated cosmic expansion. 

It is an exciting time for cluster cosmology, with various ongoing and upcoming astrophysical surveys expected to generate vast cluster catalogues, which will be used to make high-precision cosmological constraints. These catalogues will be created using all available methods of cluster detection, such as clustering of galaxies in galaxy surveys \citep[e.g.,][]{ukidss,desi,euclid,lsst}; X-ray peaks created by the hot intra-cluster gas \citep[e.g.,][]{xmm-newton,chandra,erosita}; and secondary anisotropies of the cosmic microwave background produced by the Sunyaev-Zel'dovich (SZ) effect \citep[e.g.,][]{Planck_SZ_cluster,act}.

To complement this wealth of high-quality observational data, great advances have recently been made in numerical cosmology. By incorporating sub-grid models for complex baryonic processes such as star formation, cooling, and black hole and stellar feedback \citep[e.g.,][]{2014Natur.509..177V, Schaye:2014tpa,2017MNRAS.465.3291W,Pillepich:2017jle}, it has become possible to simulate halo populations whose stellar and gaseous properties closely match those of galaxies and clusters in the real Universe. It is vital that we make use of these advances to develop robust theoretical predictions which, when combined with observational data, can generate unbiased constraints. For example, in some modified gravity (MG) models \citep[see, e.g.,][for a review]{Koyama:2015vza}, the strength of gravity is altered on large scales. While this can create observational signatures in the abundance of galaxy clusters, which can be used to make constraints, it can also alter the internal properties of clusters, such as the density profile, the mass and the thermal properties. If these effects are not studied in detail, using simulations that incorporate both full baryonic physics\footnote{Throughout this work, we will refer to simulations that include a sub-grid treatment of baryonic processes (such as cooling, star formation, and black hole feedback) as `full-physics' simulations, since these are among the most detailed and complete baryonic models that are currently available. However, we note that these treatments do not provide a complete physical description of the underlying processes.} and the MG theory of interest, the inferred constraints could be biased. 

A vital ingredient for cluster cosmology is scaling relations, which relate the cluster mass to observables including the gas temperature, the Compton $Y$-parameter of the SZ effect ($Y_{SZ}$), and the X-ray luminosity ($L_{\rm X}$). In the standard $\Lambda$CDM model, these observables form power-law relations with the mass which can be used, for instance, to relate theoretical predictions of the halo mass function (HMF), ${\rm d}n/{\rm d}\log M$, to the observable mass function, ${\rm d}n/{\rm d}Y_{\rm obs}$, defined in terms of some observable $Y_{\rm obs}$. The modelling of these scaling relations is the focus of many works, both observational \citep[e.g.,][]{Ade:2013lmv} and theoretical \citep[e.g.,][]{Truong:2016egq}. However, the presence of strengthened gravitational forces in MG models can affect the gas temperature of clusters, which in turn affects cluster observables including the examples given above. Consequently, the observable-mass scaling relations become biased, and may not even behave as power laws \citep[e.g.,][]{arnold:2014, He:2015mva}.

Another important property is the halo concentration, which is a parameter of the universal Navarro-Frenk-White \citep[NFW,][]{NFW} density profile of dark matter haloes. If the concentration can be predicted as a function of the cluster mass and redshift \citep[e.g.,][]{Duffy:2008pz,Ludlow:2013vxa,Dutton:2014xda}, this allows the cluster density profile to be modelled. This is required for conversions between different mass definitions, which is necessary if, for example, the theoretical HMF prediction and the cluster observables are defined using different spherical overdensities. The concentration also has wider uses, including the theoretical modelling of the nonlinear matter power spectrum \citep[e.g.,][]{Brax:2013fna,Lombriser:2013eza,2016PhRvD..93j3522A,Hu:2017aei,2019MNRAS.488.2121C}, which, like the HMF, can also be used to probe MG theories. It is therefore important to understand the effects that a strengthened gravitational force can have on the density profiles and concentrations of haloes.

A popular example of MG theories with a strengthened gravity is the Dvali-Gabadadze-Porrati model \citep{DVALI2000208}. This consists of two branches: the `self-accelerating' (sDGP) branch and the `normal' (nDGP) branch. The former is able to give rise to the late-time accelerated expansion without requiring an additional dark energy component; however, it is also prone to ghost instabilities \citep[e.g.,][]{Koyama:2007za}, which are absent in the latter. As a result, the nDGP branch has become the more popular model of gravity, despite requiring some additional dark energy. This model gives rise to departures from General Relativity (GR) above a particular `cross-over' scale, resulting in a `fifth force' which enhances the total strength of gravity. At smaller scales, the fifth force is screened out by the Vainshtein screening mechanism \citep{VAINSHTEIN1972393}, which ensures that the model is still consistent with, for example, Solar System tests \citep[e.g.,][]{Will:2014kxa}. The large-scale force enhancement produces observational signatures in large scale structure, and in recent years the model has been studied and tested using various probes: with cluster number counts \citep[e.g.,][]{2009PhRvD..80l3003S,vonBraun-Bates:2018lxq}, redshift-space distortions \citep[e.g.,][]{PhysRevD.94.084022,Hernandez-Aguayo:2018oxg}, the SZ angular power spectrum \citep{Mitchell:2020fnj} and cosmic voids \citep[e.g.,][]{Falck:2017rvl,10.1093/mnras/stz022}. Models which feature Vainshtein screening have also been tested by, for example, comparing weak lensing data with SZ and X-ray cluster observations \citep{Terukina:2015jua}.

However, the fifth force of nDGP could also alter cluster properties such as the temperature and density profile. If these are not taken into account, then cluster mass measurements could become biased, affecting constraints. In this work, we address this issue by studying four models of nDGP, which exhibit different strengths of the fifth force, using a combination of dark-matter-only (DMO) and full-physics simulations that cover a wide range of resolutions and box sizes. This allows us to study and model the effects of the nDGP fifth force on the halo concentration and observable-mass scaling relations. By combining our DMO simulations, we also examine the halo abundance over a continuous mass range extending from Milky Way galaxy-sized to large cluster-sized haloes.

This study forms part of a series of works aiming to develop a general framework for unbiased tests of gravity using galaxy clusters \citep[see,][for details]{Mitchell:2018qrg}. So far, our series has focused on the popular Hu-Sawicki \citep[HS,][]{Hu:2007nk} model of $f(R)$ gravity. Using a large suite of DMO simulations, we have calibrated simple yet powerful models for the enhancement of the dynamical mass \citep{Mitchell:2018qrg} and the halo concentration \citep{Mitchell:2019qke}. Most recently, we used the first simulations to simultaneously incorporate both full physics and $f(R)$ gravity to study the effects on observable-mass scaling relations \citep{Mitchell:2020aep}. Using the results of these works, we are now developing a pipeline for constraining the strength of the present-day background scalar field, $f_{R0}$, of $f(R)$ gravity using Markov chain Monte Carlo (MCMC) techniques, which we will present in an upcoming work. With the results of the present paper, we hope to put together a similar pipeline for unbiased constraints of nDGP.

This paper is organised as follows: in Sec.~\ref{sec:theory}, we briefly outline the underlying theory of the nDGP model; in Sec.~\ref{sec:methods}, we describe the nDGP simulations used in the analyses of this work and the method for calculating the halo properties; our main results are presented and discussed in Sec.~\ref{sec:results}; and, in Sec.~\ref{sec:conclusions}, we give the main conclusions and discuss the significance of our results.

\section{Theory}
\label{sec:theory}

In the nDGP model \citep{DVALI2000208}, the Universe is assumed to be a 4-dimensional brane embedded within a 5-dimensional bulk spacetime. The gravitational action is given by:
\begin{equation}
    S=\int_{\rm brane} {\rm d}^4x\sqrt{-g}\left(\frac{R}{16\pi G}\right) + \int {\rm d}^5x\sqrt{-g^{(5)}}\left(\frac{R^{(5)}}{16\pi G^{(5)}}\right).
\label{eq:DGP_action}
\end{equation}
The first integral represents the contribution from the 4-dimensional brane. This is equivalent to the Einstein-Hilbert action of GR, where $R$ is the Ricci scalar curvature, $G$ is Newton's gravitational constant and $g$ is the determinant of the metric tensor $g_{\alpha\beta}$ (Greek indices run over $0,1,2,3$). The second integral represents the contribution from the 5-dimensional bulk, where $R^{(5)}$, $G^{(5)}$ and $g^{(5)}$ are analogous to $R$, $G$ and $g$.

The ratio of the gravitational constants, $G^{(5)}/G$, defines a characteristic scale which is known as the `cross-over' scale $r_{\rm c}$:
\begin{equation}
    r_{\rm c} = \frac{1}{2}\frac{G^{(5)}}{G}.
\label{eq:DGP_crossover}
\end{equation}
Above the cross-over scale, the second term of Eq.~(\ref{eq:DGP_action}) dominates and the behaviour of gravity diverges from GR. The cross-over scale is often re-expressed using the dimensionless parameter $\Omega_{\rm rc}$, which is given by:
\begin{equation}
    \Omega_{\rm rc} \equiv \frac{1}{4H_0^2r_{\rm c}^2},
\label{eq:DGP_omega}
\end{equation}
where $H_0$ is the Hubble constant. Deviations from GR are typically characterised using the quantity $H_0r_{\rm c}$. In this work, we study models with $H_0r_{\rm c}$ equal to 5, 2, 1 and 0.5, and we will refer to these as N5, N2, N1 and N0.5, respectively. 

Assuming that the background is homogeneous and isotropic, the time evolution of the Hubble parameter is given by:
\begin{equation}
    \frac{H(a)}{H_0} = \sqrt{\Omega_{\rm M}a^{-3} + \Omega_{\rm DE}(a) + \Omega_{\rm rc}} - \sqrt{\Omega_{\rm rc}},
\label{eq:DGP_friedmann}
\end{equation}
where $\Omega_{\rm M}$ is the present-day dimensionless matter density, $a$ is the cosmic scale factor and $\Omega_{\rm DE}(a)$ is the dimensionless density of the dark energy component, which is included in nDGP to ensure that $H(a)$ is consistent with the background expansion history that we observe. We assume that $H(a)$ matches that of a flat $\Lambda$CDM cosmology with the same $\Omega_{\rm M}$, and that the clustering of this dark energy component is negligible on the sub-horizon scales that we are interested in.

Structure formation in nDGP is governed by the modified Poisson equation, which, in the weak-field and quasi-static limits, is given by \Cesar{\citep{Koyama:2007ih}}:
\begin{equation}
    \nabla^2\Phi = 4\pi Ga^2\delta\rho_{\rm M} + \frac{1}{2}\nabla^2\varphi,
\label{eq:DGP_pot}
\end{equation}
where $\Phi$ is the Newtonian gravitational potential, $\delta \rho_{\rm M}$ represents the perturbations in the matter density field, and $\varphi$ is an additional scalar field which describes the position of the brane in the 5D bulk, known as the brane-bending mode, and which is the new degree of freedom of the DGP model. Departures from GR are encapsulated in the scalar field term, which obeys the following dynamical equation of motion \Cesar{\citep{Koyama:2007ih}}:
\begin{equation}
    \nabla^2\varphi + \frac{r_{\rm c}^2}{3\beta a^2}\left[(\nabla^2\varphi)^2 - (\nabla_i\nabla_j\varphi)(\nabla^i\nabla^j\varphi)\right] = \frac{8\pi Ga^2}{3\beta}\delta\rho_{\rm M},
\label{eq:DGP_scalar_field}
\end{equation}
The time-dependent function $\beta$ is given by:
\begin{equation}
    \beta(a) = 1 + 2Hr_{\rm c}\left(1 + \frac{\dot{H}}{3H^2}\right) = 1 + \frac{\Omega_{\rm M}a^{-3} + 2\Omega_{\Lambda}}{2\sqrt{\Omega_{\rm rc}\left(\Omega_{\rm M}a^{-3} + \Omega_{\Lambda}\right)}},
\label{eq:DGP_beta}
\end{equation}
where $\Omega_{\Lambda}\equiv1-\Omega_{\rm M}$. The nonlinear terms in the square bracket of Eq.~(\ref{eq:DGP_scalar_field}) are negligible on sufficiently large and linear scales, giving rise to an additional `fifth force' which enhances the total strength of gravity by factor $[1 + 1/(3\beta)]$. The fifth force is stronger at later times: at the present-day, the fifth force enhances the total strength of gravity by factors $1.04$, $1.08$, $1.12$ and $1.18$ in N5, N2, N1 and N0.5, respectively. The screening %out 
of the fifth force on small scales, where the nonlinear terms in Eq.~(\ref{eq:DGP_scalar_field}) cannot be ignored, is known as the Vainshtein screening mechanism \citep{VAINSHTEIN1972393}, which is very efficient at suppressing the fifth force inside and near massive astrophysical objects.

\section{Simulations and methods}
\label{sec:methods}

Since Eq.~\eqref{eq:DGP_scalar_field} is highly nonlinear, the fifth force in the nDGP model can display a wide spectrum of behaviours, depending on time, scale and mass of the objects being considered. Therefore, numerical simulations are essential for predicting its cosmological properties and implications accurately. For earlier works that make use of nDGP simulations, see, e.g., \citet{Chan:2009ew,Schmidt:2009sg,Khoury:2009tk,Li:2013nua,Falck:2014jwa,Falck:2015rsa}. We describe the DMO and full-physics simulations used in this work in Sec.~\ref{sec:methods:simulations}. Then, in Sec.~\ref{sec:methods:measurements}, we explain our methods for computing the thermal properties and concentration of our haloes.

\subsection{Simulations}
\label{sec:methods:simulations}

Our simulations were run using the \textsc{arepo} code \citep{2010MNRAS.401..791S}, which can be used to run $N$-body and hydrodynamical cosmological simulations. The code includes a sub-grid treatment of full baryonic physics, including star formation, cooling, and stellar and black hole feedback, which is implemented using the IllustrisTNG model \citep[for a complete description, see][]{2017MNRAS.465.3291W,Pillepich:2017jle}. The code also features a new MG solver, which uses adaptive mesh refinement techniques to calculate the highly nonlinear fifth force in MG models including HS $f(R)$ gravity \citep{Arnold:2019vpg} and nDGP \citep{Hernandez-Aguayo:2020kgq}. 

%%%%%%%%%%%%%%%%%%%%%%%%%%%%%%%%%%%%%
\begin{table*}
\centering

\small
\begin{tabular}{ c@{\hskip 0.5in}ccccc } 
 \toprule
 
 Specifications & \multicolumn{5}{c}{Simulations} \\
 and models & \textsc{shybone} & L62 & L200 & L500 & L1000 \\

 \midrule

 box size / $h^{-1}$Mpc & 62 & 62 & 200 & 500 & 1000 \\ 
 particle number & $2\times(512^3)$ & $512^3$ & $1024^3$ & $1024^3$ & $1024^3$ \\ 
 DM particle mass / $h^{-1}M_{\odot}$ & $1.28\times10^8$ & $1.52\times10^8$ & $6.39\times10^8$ & $9.98\times10^9$ & $7.98\times10^{10}$ \\
 gas particle mass / $h^{-1}M_{\odot}$ & $\approx2.5\times10^7$ & - & - & - & - \\
 nDGP models & N1, N5 & N1, N5 & N0.5, N1, N2, N5 & N0.5, N1, N2, N5 & N0.5, N1, N2, N5\\
 
 \bottomrule
 
\end{tabular}

\caption{Specifications of the \textsc{arepo} simulations used in this investigation. The four dark-matter-only simulations are labelled L62, L200, L500 and L1000, according to their box size. The simulations have all been run for GR in addition to the nDGP models listed, where N0.5, N1, N2 and N5 correspond to $H_0r_{\rm c}=0.5, 1, 2, 5$, respectively.}
\label{table:simulations}

\end{table*}
%%%%%%%%%%%%%%%%%%%%%%%%%%%%%%%%%%%%%

The specifications of our simulations are provided in Table~\ref{table:simulations}. One of these is the first cosmological simulation to simultaneously incorporate both full baryonic physics and nDGP\footnote{We note that the IllustrisTNG model was tuned using standard gravity simulations. However, the differences between the GR and nDGP predictions for the stellar and gas properties of galaxies are generally small compared to typical observational scatters \citep[see, e.g., Fig.~8 of][]{Hernandez-Aguayo:2020kgq}, making a full retuning of the TNG parameters for the nDGP model unnecessary.}. This simulation, which is part of the \textsc{shybone} simulation suite \citep[see][]{Arnold:2019vpg,Hernandez-Aguayo:2020kgq}, has box size $62h^{-1}{\rm Mpc}$ and consists of $512^3$ dark matter particles, with mass $1.28\times10^8h^{-1}M_{\odot}$, and (initially) the same number of Voronoi gas cells, which have mass $\sim2.5\times10^7h^{-1}M_{\odot}$ on average. We also have four DMO $N$-body simulations, with box sizes $62h^{-1}{\rm Mpc}$, $200h^{-1}{\rm Mpc}$, $500h^{-1}{\rm Mpc}$ and $1000h^{-1}{\rm Mpc}$. Throughout this work, we refer to these as L62, L200, L500 and L1000, respectively. These span a wide range of mass resolutions -- from $1.52\times10^8h^{-1}M_{\odot}$ in L62 to $7.98\times10^{10}h^{-1}M_{\odot}$ in L1000 -- allowing us to study haloes spanning, continuously, the mass range $\sim10^{11}h^{-1}M_{\odot}$ to $\sim10^{15}h^{-1}M_{\odot}$. 

The simulations have all been run with cosmological parameters $(h,\Omega_{\rm M},\Omega_{\rm b},\sigma_8,n_{\rm s})$=$(0.6774,0.3089,0.0486,0.8159,0.9667)$, where $h=H_0/(100{\rm kms^{-1}Mpc^{-1}})$, $\Omega_{\rm b}$ is the dimensionless baryonic density parameter, $\sigma_8$ is the present-day linear fluctuation of the density field at the scale $8h^{-1}{\rm Mpc}$, and $n_{\rm s}$ is the slope of the primordial matter power spectrum. All simulations include runs with N5 and N1, in addition to GR. The L200, L500 and L1000 simulations also feature runs with N2 and N0.5, allowing us to thoroughly explore the effects of different strengths of the fifth force on halo properties. The simulations all begin at redshift $z=127$. For this work, we use 12 particle snapshots from each simulation which span the redshift range $0\leq z\leq3$.

%%%%%%%%%%%%%%%%%%%%%%%%%%%%%%%%%%%%%%%%%%%%%%%%%
\begin{figure*}
\centering
\includegraphics[width=1.0\textwidth]{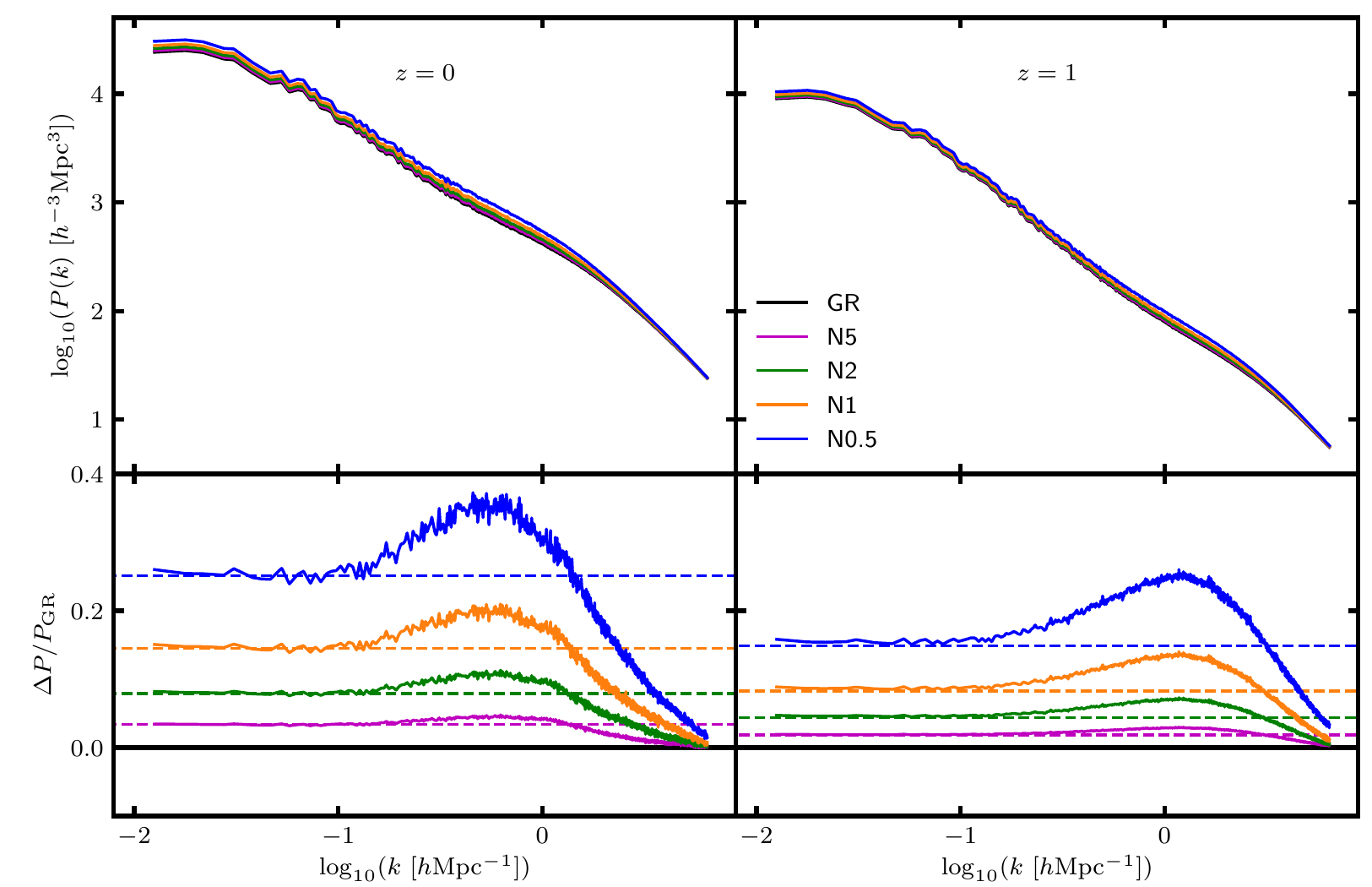}
\caption{[{\it Colour Online}] Matter power spectrum (\textit{top row}) and its relative difference in nDGP with respect to GR (\textit{bottom row}), as a function of the wavenumber at redshifts $0$ and $1$. The data has been generated using our dark-matter-only L500 simulation (see Table \ref{table:simulations}), which has been run for GR (\textit{black}) and the nDGP models N5 (\textit{magenta}), N2 (\textit{green}), N1 (\textit{orange}) and N0.5 (\textit{blue}). The dashed lines in the bottom row show the linear theory predictions of the relative difference.}
\label{fig:Pk}
\end{figure*}
%%%%%%%%%%%%%%%%%%%%%%%%%%%%%%%%%%%%%%%%%%%%%%%%%%

For completeness and as a first check, we show here the matter power spectra of the nDGP models simulated in this work; because this is not the primary focus, we shall only discuss the result briefly. The upper panels of Fig.~\ref{fig:Pk} show the matter power spectra generated using the $z=0$ and $z=1$ snapshots of L500 (similar results can be found for the L200 and L1000 boxes). The relative differences between the nDGP and GR spectra are shown in the lower panels, where we have also included the predictions from linear theory (dashed lines). On large scales ($k\lesssim0.1h{\rm Mpc}^{-1}$), the observed relative differences closely match the linear predictions; here, the fifth force enhances the power by $\sim25\%$ in N0.5 and by a few percent in N5 at $z=0$. The enhancement is even greater at intermediate scales, where the N0.5 power is enhanced by up to $\sim35\%$. This is a consequence of mode-coupling at these scales. At smaller scales ($k\gtrsim1h{\rm Mpc}^{-1}$), which correspond to halo scales, the Vainshtein screening of the fifth force suppresses the power spectrum enhancement. These results are consistent with previous works \citep[e.g.,][]{2009PhRvD..80l3003S,2010PhRvD..81f3005S,Winther:2015wla}. While the trends are similar at $z=0$ and $z=1$, the nDGP enhancement is smaller for the latter due to the fifth force being weaker at earlier times.

\subsection{Halo catalogues}
\label{sec:methods:measurements}

At each particle snapshot, we have generated halo catalogues using the \textsc{subfind} code \citep{springel2001}, which uses the friends-of-friends (FOF) algorithm and gravitational unbinding to locate FOF groups and their underlying substructure. We will refer to FOF groups as `haloes' throughout this work. The halo mass, $M_{\Delta}$, is defined as the total mass (for hydrodynamical runs this includes the mass of gas, star particles and black holes) that is enclosed within the sphere that is centred on the position of the most bound particle and contains an average density of $\Delta$ times the critical density of the Universe at the halo redshift. The halo radius, $R_{\Delta}$, is the radius of this sphere. In this work, we have considered overdensities $\Delta=200$ and $\Delta=500$, which correspond to masses $M_{200}$ and $M_{500}$, respectively.

\subsubsection{Gas and thermal properties}

We have calculated the thermal properties of our haloes using gas particles found within the radius $R_{500}$. The temperature of each gas particle is computed using the internal energy and electron abundance, which are outputted by \textsc{arepo}. Here, we assume that the adiabatic index is equal to $5/3$ and that the primordial hydrogen mass fraction is equal to $0.76$. We then evaluate the mass-weighted gas temperature:
\begin{equation}
    \bar{T}_{\rm gas} = \sum_i\frac{m_{{\rm gas},i}T_i}{m_i},
\end{equation}
where $m_{{\rm gas},i}$ and $T_i$ are the mass and temperature of gas cell $i$, respectively, and the summation carries over all particles in the range $0.15R_{500}<r<R_{500}$. This excludes the core region, which we define as the radial range $r<0.15R_{500}$, where dynamical (e.g., mergers) and thermal (e.g., feedback) processes can cause significant dispersion in the temperature profile. This exclusion is consistent with previous works that use simulations to study observable-mass scaling relations \citep[e.g.,][]{Fabjan:2011,Brun:2016jtk,Truong:2016egq}. In \citet{Mitchell:2020aep}, where we used the $f(R)$ subset of the \textsc{shybone} simulations to study scaling relations in $f(R)$ gravity and GR, we also considered core regions $r<0.1R_{500}$ and $r<0.2R_{500}$, and found that the effect of the exclusion radius on the model differences is negligible.

In addition to the gas temperature, we also study the Compton $Y$-parameter of the SZ effect, $Y_{\rm SZ}$, and its X-ray analogue, $Y_{\rm X}$. The $Y_{\rm SZ}$ parameter is a measure of the integrated SZ flux:
\begin{equation}
    Y_{\rm SZ} = \frac{\sigma_{\rm T}}{m_{\rm e}c^2}\sum_iN_{{\rm e},i}T_i,
\end{equation}
where $\sigma_{\rm T}$ is the Thomson electron scattering cross section, $m_{\rm e}$ is the electron rest mass, $c$ is the speed of light in a vacuum and $N_{{\rm e},i}$ is the number of electrons in gas cell $i$. The summation again runs over the radial range described above. The $Y_{\rm X}$ parameter is given by the product of the gas mass and the mass-weighted temperature:
\begin{equation}
    Y_{\rm X} = M_{\rm gas}\times\bar{T}_{\rm gas},
\end{equation}
where $M_{\rm gas}$ is the total mass of all gas particles within the radial range $r<R_{500}$ (including the core region).

\subsubsection{Halo concentration}
\label{sec:methods:concentration}

The halo concentration is a parameter of the NFW profile \citep{NFW}:
\begin{equation}
    \rho(r) = \frac{\rho_{\rm s}}{(r/R_{\rm s})(1 + r/R_{\rm s})^2},
    \label{eq:nfw_profile}
\end{equation}
where $\rho_{\rm s}$ is the characteristic density and $R_{\rm s}$ is the scale radius, which is the radius at which the profile transitions from an $r^{-1}$ power law (inner regions) to an $r^{-3}$ power law (outer regions). The concentration is defined as $c_{200}=R_{200}/R_{\rm s}$.\footnote{In literature, the concentration is usually defined with respect to overdensity $200$, so this is the definition that we focus on in this work. However, as long as the concentration $c_{\Delta}$ is known for some overdensity $\Delta$, then the value can be inferred for any other overdensity.} If both the mass (or radius) and concentration of a halo are known, then it is straightforward to calculate the scale radius and the characteristic density.

We have measured the concentration by fitting the NFW profile to the density profiles of individual haloes. To do this, we first rewrite Eq.~(\ref{eq:nfw_profile}) in terms of the dimensionless radial distance $x=r/R_{200}$ and take the logarithm of both sides:
\begin{equation}
    \log_{10}\rho = \log_{10}\rho_{\rm s} - \log_{10}(xc_{200}) - 2\log_{10}(1 + xc_{200}).
    \label{eq:nfw_log}
\end{equation}
We measure the halo density within 20 radial bins, which are equally spaced in $\log(x)$, from $x=0.05$ to $x=1$ ($r=R_{200}$). Logarithmic bins are used so that the inner regions and outer regions are equally-well fitted. The radial range also excludes the innermost and outermost regions where the halo may be poorly resolved and the density underestimated. We used unweighted least squares to fit Eq.~(\ref{eq:nfw_log}) to the density profile by varying $\rho_{\rm s}$ and $c_{200}$. The concentration of the halo is given by the best-fit value of $c_{200}$.

Various alternative approaches, involving relations between the concentration and the maximum circular velocity of a halo, have also been described in the literature \citep[e.g.,][]{Springel:2008cc,Prada:2011jf}. These relations are approximations which have been derived by assuming that the halo density profile behaves according to Eq.~(\ref{eq:nfw_profile}). However, this can lead to biased measurements of the concentration if the halo density does not perfectly follow the NFW profile. For example, in \citet{Mitchell:2019qke}, we found that, for haloes in HS $f(R)$ gravity (where the density is often enhanced at the inner halo regions), the concentration can be overestimated when the relation from \citet{Springel:2008cc} is used. The concentration is, by definition, a parameter of the NFW profile, and so a full fitting of Eq.~(\ref{eq:nfw_log}) to the halo density profile is the most reliable option.

\section{Results}
\label{sec:results}

In Sec.~\ref{sec:results:scaling_relations}, we present our results for the observable-mass scalings using our full-physics simulations. Then, in Sec.~\ref{sec:results:concentration}, we study and model the concentration-mass-redshift relation in nDGP. Finally, in Sec.~\ref{sec:results:hmf}, we examine the HMF in nDGP.

\subsection{Observable-mass scaling relations}
\label{sec:results:scaling_relations}

%%%%%%%%%%%%%%%%%%%%%%%%%%%%%%%%%%%%%%%%%%%%%%%%%
\begin{figure}
\centering
\includegraphics[width=\columnwidth]{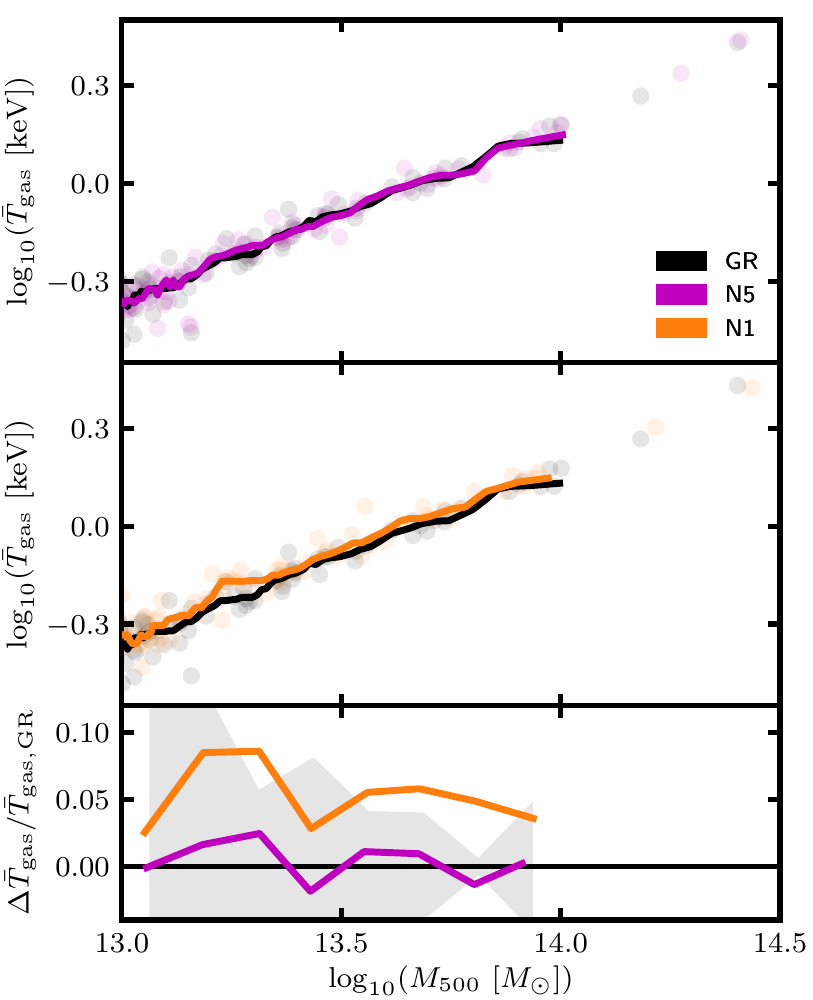}
\caption{[{\it Colour Online}] Gas temperature as a function of mass for haloes from our full-physics \textsc{shybone} simulations (see Sec.~\ref{sec:methods:simulations}) at $z=0$. Data is included for GR (\textit{black}) and the nDGP models N5 (\textit{magenta}) and N1 (\textit{orange}). The data points correspond to individual haloes. The lines show the median temperature and mean logarithm of the mass which have been computed using a moving window. \textit{Bottom panel}: relative difference between the median temperatures in nDGP and GR; the grey shaded region shows the size of the GR halo scatter.}
\label{fig:tgas}
\end{figure}
%%%%%%%%%%%%%%%%%%%%%%%%%%%%%%%%%%%%%%%%%%%%%%%%%%

%%%%%%%%%%%%%%%%%%%%%%%%%%%%%%%%%%%%%%%%%%%%%%%%%
\begin{figure}
\centering
\includegraphics[width=\columnwidth]{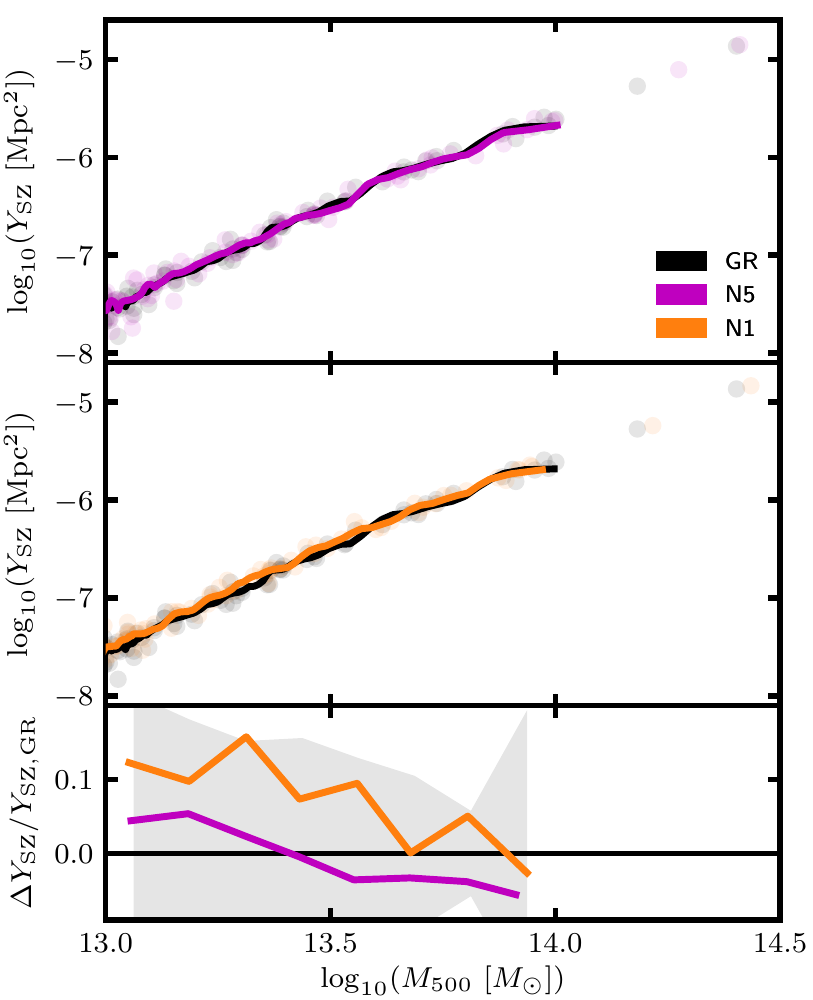}
\caption{[{\it Colour Online}] SZ Compton $Y$-parameter as a function of mass for haloes from our full-physics \textsc{shybone} simulations (see Sec.~\ref{sec:methods:simulations}) at $z=0$. Data is included for GR (\textit{black}) and the nDGP models N5 (\textit{magenta}) and N1 (\textit{orange}). The data points correspond to individual haloes. The lines show the median $Y$-parameter and mean logarithm of the mass which have been computed using a moving window. \textit{Bottom panel}: relative difference between the median $Y$-parameters in nDGP and GR; the grey shaded region shows the size of the GR halo scatter.}
\label{fig:ysz}
\end{figure}
%%%%%%%%%%%%%%%%%%%%%%%%%%%%%%%%%%%%%%%%%%%%%%%%%%

%%%%%%%%%%%%%%%%%%%%%%%%%%%%%%%%%%%%%%%%%%%%%%%%%
\begin{figure}
\centering
\includegraphics[width=\columnwidth]{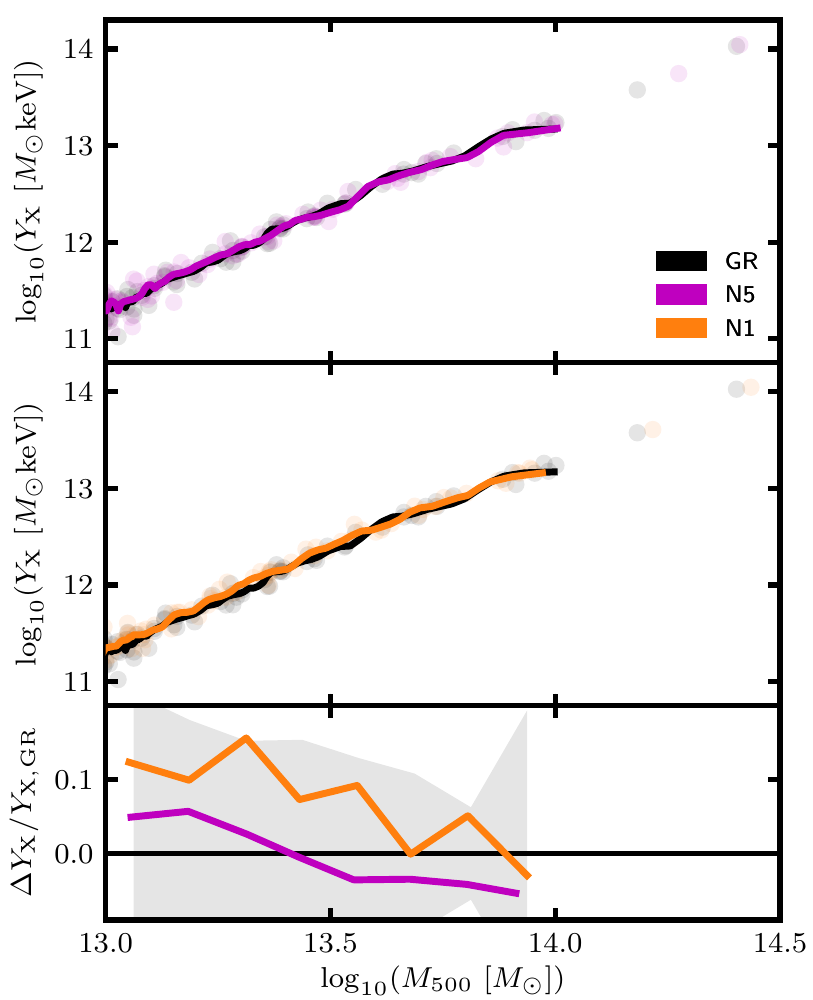}
\caption{[{\it Colour Online}] X-ray analogue of the Compton $Y$-parameter as a function of mass for haloes from our full-physics \textsc{shybone} simulations (see Sec.~\ref{sec:methods:simulations}) at $z=0$. Data is included for GR (\textit{black}) and the nDGP models N5 (\textit{magenta}) and N1 (\textit{orange}). The data points correspond to individual haloes. The lines show the median $Y$-parameter and mean logarithm of the mass which have been computed using a moving window. \textit{Bottom panel}: relative difference between the median $Y$-parameters in nDGP and GR; the grey shaded region shows the size of the GR halo scatter.}
\label{fig:yx}
\end{figure}
%%%%%%%%%%%%%%%%%%%%%%%%%%%%%%%%%%%%%%%%%%%%%%%%%%

%%%%%%%%%%%%%%%%%%%%%%%%%%%%%%%%%%%%%%%%%%%%%%%%%
\begin{figure*}
\centering
\includegraphics[width=0.8\textwidth]{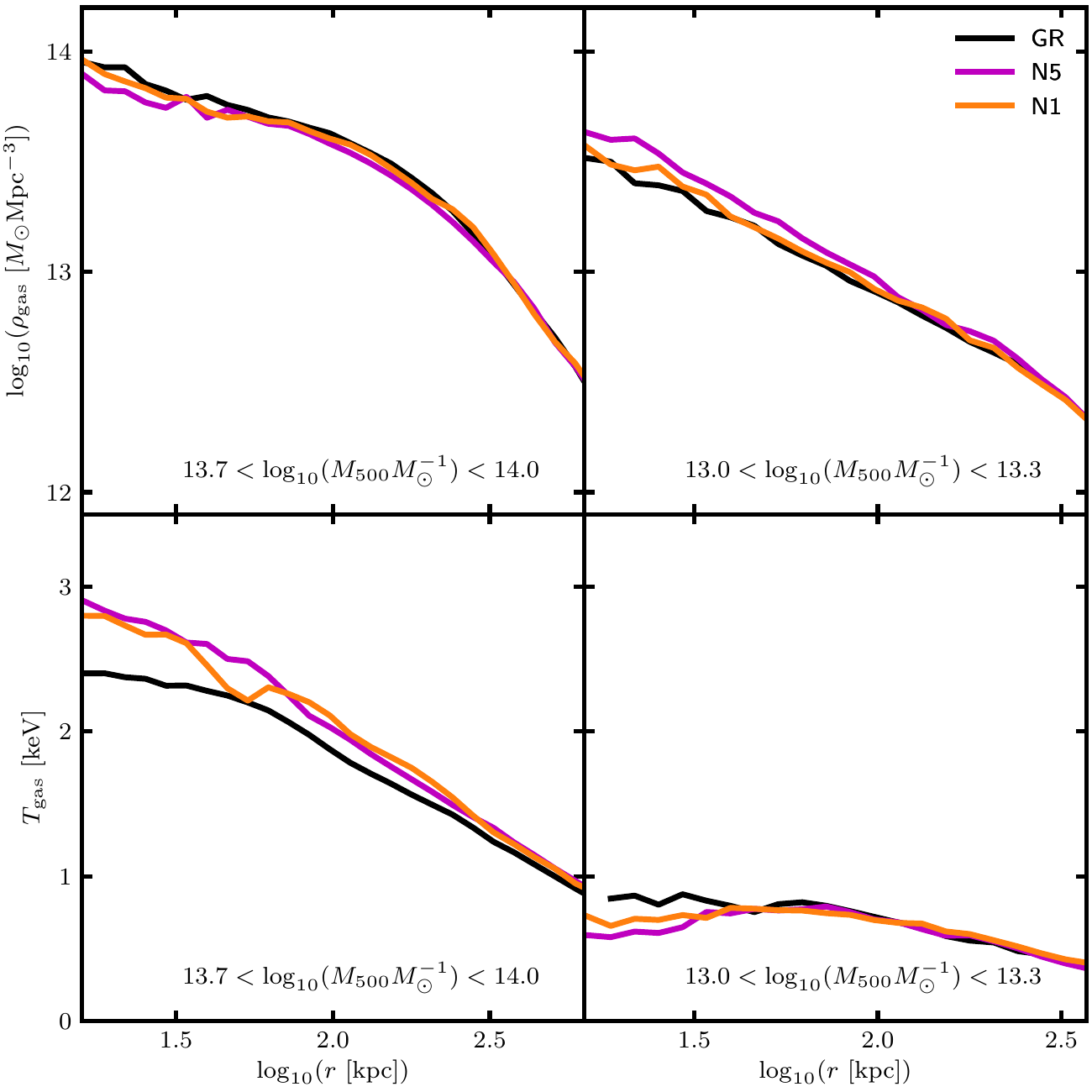}
\caption{[{\it Colour Online}] Median gas density (\textit{top row}) and temperature (\textit{bottom row}) profiles of haloes from our full-physics \textsc{shybone} simulations (see Sec.~\ref{sec:methods:simulations}) at $z=0$. Data is shown for GR (\textit{black}) and the nDGP models N5 (\textit{magenta}) and N1 (\textit{orange}). The two mass bins used to measure the median profiles are annotated. The maximum radius shown for each column corresponds to $R_{500}$.}
\label{fig:profiles}
\end{figure*}
%%%%%%%%%%%%%%%%%%%%%%%%%%%%%%%%%%%%%%%%%%%%%%%%%%

In Figs.~\ref{fig:tgas}-\ref{fig:yx}, we plot the mass-weighted gas temperature and the $Y_{\rm SZ}$ and $Y_{\rm X}$ parameters against the halo mass $M_{500}$. In addition to showing individual data points for each halo in the mass range $M_{500}>10^{13}M_{\odot}$, we also plot lines showing the median observable as a function of the mean logarithm of the mass. These averages have been computed using a moving window with a fixed size of 10 haloes. This approach, which is consistent with our study of the observable-mass scaling relations using the $f(R)$ \textsc{shybone} simulations \citep{Mitchell:2020aep}, is preferred over using a set of fixed-width bins, which would contain much fewer haloes at high mass than at low mass. The moving averages make use of all haloes with mass $M_{500}>10^{13}M_{\odot}$, including cluster-sized haloes with $M_{500}\gtrsim10^{14}M_{\odot}$. We note, however, that because there are only a few haloes with this mass (owing to the small box size of the full-physics simulations), the highest mean mass of the moving average is only $\sim10^{14}M_{\odot}$. The lower panels of Figs.~\ref{fig:tgas}-\ref{fig:yx} show the relative differences between the observable medians in nDGP and GR. These are smoothed by computing the mean relative difference within 8 mass bins. We also show the root-mean-square halo scatter in GR for each of these bins (grey shaded regions). 

The $\bar{T}_{\rm gas}$-$M$ relation is shown in the top two panels of Fig.~\ref{fig:tgas}. Both the GR and nDGP data follow a power-law relation as a function of the mass. This behaviour is a result of the intrinsic connection between the gravitational potential and thermal properties \citep[e.g.,][]{1986MNRAS.222..323K,Voit:2004ah}: during the formation of groups and clusters of galaxies, the initial gravitational potential energy of in-falling gas is converted to thermal energy during shock-heating. This produces the intrinsic scaling relation shown in the figure, with higher-mass objects having a higher gas temperature, and this is also the reason that other thermal observables, like $Y_{\rm SZ}$ (Fig.~\ref{fig:ysz}) and $Y_{\rm X}$ (Fig.~\ref{fig:yx}) show similar power-law relations as a function of the mass.

From the lower panel of Fig.~\ref{fig:tgas}, we see that the median temperature in N5 agrees very closely with GR, typically within a couple of percent. This is consistent with the fact that the fifth force has a very small amplitude in this model (see the discussion below Eq.~\eqref{eq:DGP_beta}). However, the temperature in N1 is enhanced by about 5\% relative to GR on average. This result is quite surprising: using the same full-physics simulations, \citet{Hernandez-Aguayo:2020kgq} found that the N1 fifth force reaches just 2\%-3\% of the strength of the Newtonian force at the radius $R_{500}$ for galaxy group-sized haloes and is even more efficiently screened at smaller radii. Therefore, the total gravitational potential at radius $R_{500}$, within which we have calculated the gas temperature, is expected to be just a few percent deeper than the Newtonian potential. From the above discussion of the connection between the gravitational potential and the thermal properties, we would therefore expect the temperature to be enhanced by just a few percent rather than the 5\% that we observe. However, we note that, in nDGP, gravity is enhanced at the outer halo regions even at redshift $z=2$ \citep[see, for example, Fig.~7 of][]{Hernandez-Aguayo:2020kgq}. Therefore, between $0<z<2$, gas at the outer halo radii will undergo a gravitational acceleration in nDGP that is enhanced compared to GR. Consequently, it will have a higher speed than in GR as it reaches smaller radii where it gets shock-heated. The fact that this happens over a long period of time can potentially explain how the gas temperature is enhanced by as much as 5\% within $R_{500}$. 

Our results for the $Y_{\rm SZ}$-$M$ and $Y_{\rm X}$-$M$ scaling relations are shown in Figs.~\ref{fig:ysz} and \ref{fig:yx}, respectively. The $Y_{\rm SZ}$ and $Y_{\rm X}$ parameters are closely related to each other, and so the results appear similar for both: the enhancement of the $Y$-parameters in the N1 model ranges from zero at high masses to 10\%-15\% at low masses, while in N5 it ranges between a 5\% suppression at high masses and 5\% enhancement at low masses. The low-mass enhancement in N1 can in part be explained by the enhanced temperature seen in Fig.~\ref{fig:tgas}. Even for N5, the temperature appears to be enhanced on average for masses $M_{500}\lesssim10^{13.4}h^{-1}M_{\odot}$, so this can also partly explain the $\sim5\%$ enhancement of the $Y$-parameters at these masses. The $Y$-parameters are also correlated with the gas density. In the top row of Fig.~\ref{fig:profiles}, we show the median gas density profiles for haloes from two mass bins (annotated). For the low-mass bin, both the N5 and N1 gas profiles appear to be enhanced, on average, with respect to GR, while for the high-mass bin the profiles appear to be suppressed. This can help explain why the $Y$-parameters are enhanced in nDGP at lower masses and closer to GR or suppressed at higher masses.

The physical origin of these effects on the gas density is not entirely clear. They could be related to the complex interrelations between the nDGP fifth force and baryonic processes such as cooling and feedback. For example, if the fifth force leads to a larger amount of feedback, this would heat up and blow out surrounding gas. This would be consistent with the results shown for the gas density and temperature profiles in the high-mass bin in Fig.~\ref{fig:profiles}, where the gas density is suppressed and the temperature is enhanced in nDGP compared to GR. The opposite trend is present in the low-mass bin, which would be consistent with a lowering of feedback efficiency in nDGP compared to GR. Another possibility is that the enhancement of the gas speeds due to the fifth force leads to differences in the density profiles between nDGP and GR. This effect can be inherited from times before the gas falls into haloes and is screened from the fifth force. Haloes of different mass will experience this effect to a different extent as larger haloes are formed from matter and gas further afield.

In the lower panels of Fig.~\ref{fig:profiles} we show the halo gas temperature profiles. For the higher mass bin, the profiles in N5 and N1 are both enhanced compared to GR. For N1, this is consistent with the result for the mass-weighted temperature discussed above; however, for N5, the enhancement relative to GR appears to contradict Fig.~\ref{fig:tgas}. This is actually related to the difference in binning: while the median mass-weighted temperature has been computed using a moving window containing a fixed number of haloes, the temperature profile is computed within a single wide bin. The mean mass of this bin is actually higher in N5 than in GR, indicating that this bin contains a greater number of high-mass haloes in N5 which also have a higher temperature. This supports our decision to use a moving average in Figs.~\ref{fig:tgas}-\ref{fig:yx}, which avoids the issues that arise from having a fixed set of bins for each model. For the lower mass bin, the nDGP temperature profiles are suppressed for radii $r\lesssim 100{\rm kpc}$ and the N1 profile is just slightly enhanced at higher radii. We note that, because there are more particles at the outer radii, which cover a larger volume, these regions have a greater overall contribution to the mass-weighted temperature, which can explain why the latter is enhanced in N1 even though the temperature profile is suppressed at lower radii compared to GR. And, as described above, the difference in binning can make it difficult to directly compare Figs.~\ref{fig:tgas} and \ref{fig:profiles}.

We finally note that, due to the small box size of our full-physics simulations, we can only rigorously study the scaling relations for halo masses corresponding to galaxy groups. A larger box will be required to rigorously probe the interplay between the fifth force and baryonic physics in galaxy clusters. Galaxy groups, particularly low-mass groups, are typically more susceptible to feedback than cluster-sized objects. This is why, in Fig.~\ref{fig:tgas}, the scatter in the GR halo temperature is above 10\% for low-mass groups and less than 5\% for high-mass groups. It will therefore be interesting to see how the nDGP scaling relations compare to GR at these larger masses, where the unpredictable effects from feedback are not as significant. We plan to address this question by running large-box full-physics simulations of the nDGP model in the future.

\subsection{Concentration-mass relation}
\label{sec:results:concentration}

%%%%%%%%%%%%%%%%%%%%%%%%%%%%%%%%%%%%%%%%%%%%%%%%%
\begin{figure*}
\centering
\includegraphics[width=1.0\textwidth]{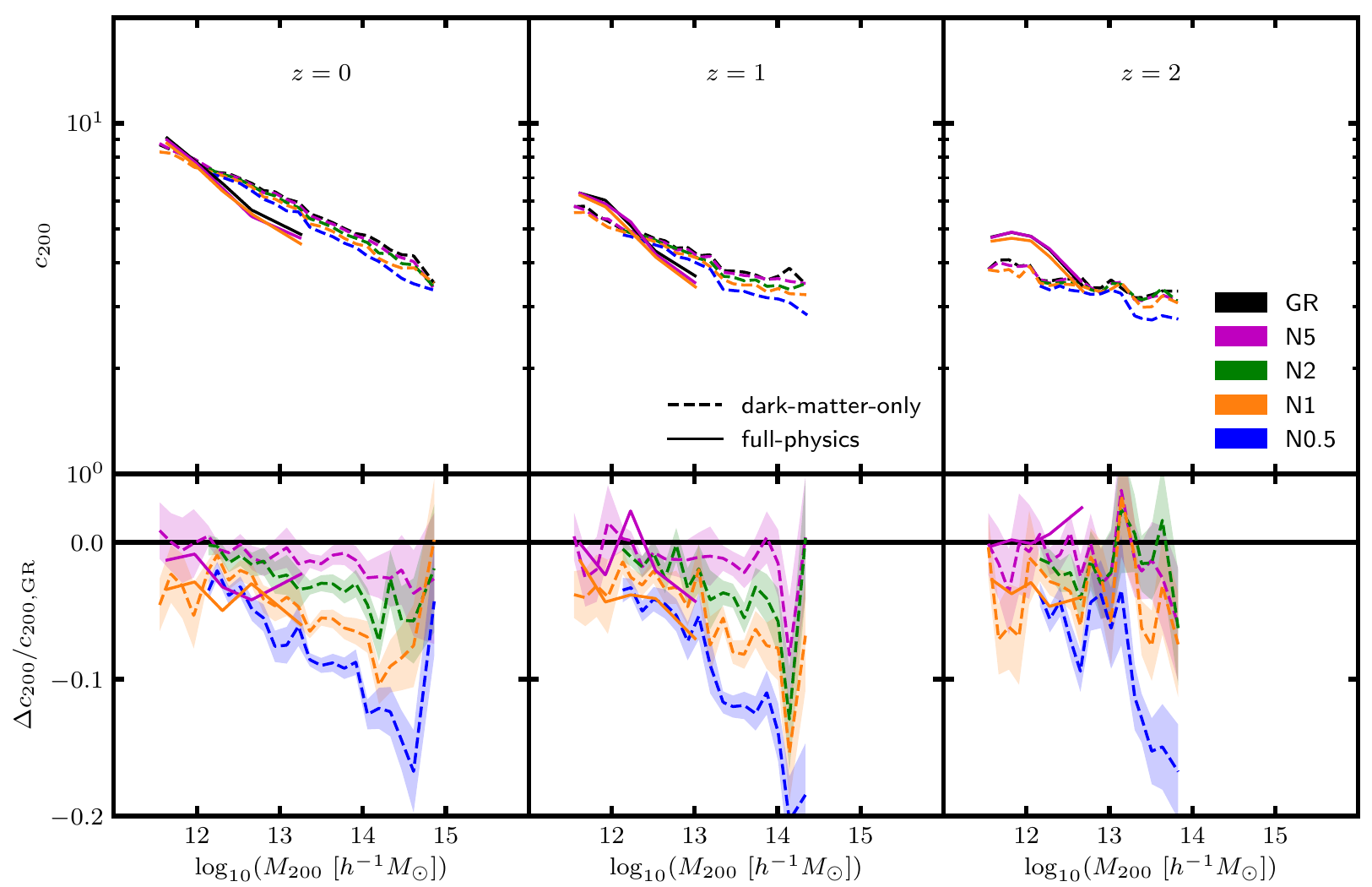}
\caption{[{\it Colour Online}] Median halo concentration (\textit{top row}) and its relative difference with respect to GR (\textit{bottom row}), as a function of the mean logarithm of the halo mass at redshifts $0$, $1$ and $2$. The data is generated using the dark-matter-only simulations L62, L200 and L500 (\textit{dashed lines}) and our full-physics simulation (\textit{solid lines}), the specifications of which are given in Table \ref{table:simulations}. These have been run for GR (\textit{black}) and the nDGP models N5 (\textit{magenta}), N2 (\textit{green}), N1 (\textit{orange}) and N0.5 (\textit{blue}). The shaded regions in the lower panels show the $1\sigma$ uncertainty in the relative difference.}
\label{fig:c_M}
\end{figure*}
%%%%%%%%%%%%%%%%%%%%%%%%%%%%%%%%%%%%%%%%%%%%%%%%%%

%%%%%%%%%%%%%%%%%%%%%%%%%%%%%%%%%%%%%%%%%%%%%%%%%
\begin{figure*}
\centering
\includegraphics[width=1.0\textwidth]{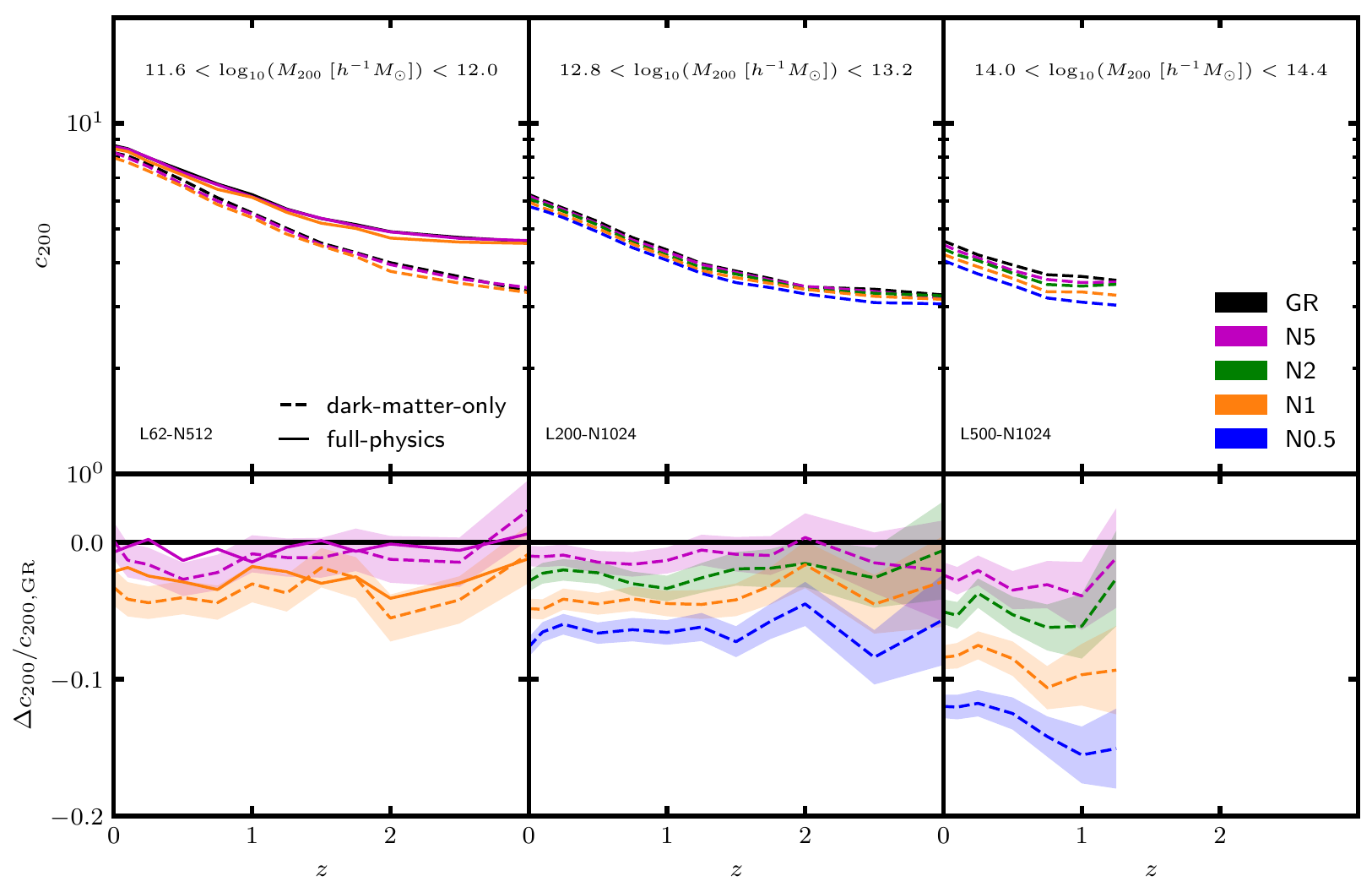}
\caption{[{\it Colour Online}] Median halo concentration (\textit{top row}) and its relative difference with respect to GR (\textit{bottom row}), as a function of redshift for three mass bins. The data is generated using the dark-matter-only simulations L62, L200 and L500 (\textit{dashed lines}) and the full-physics simulation  (\textit{solid lines}), the specifications of which are given in Table \ref{table:simulations}. These have been run for GR (\textit{black}) and the nDGP models N5 (\textit{magenta}), N2 (\textit{green}), N1 (\textit{orange}) and N0.5 (\textit{blue}). The shaded regions in the lower panels show the $1\sigma$ uncertainty in the relative difference.}
\label{fig:c_z}
\end{figure*}
%%%%%%%%%%%%%%%%%%%%%%%%%%%%%%%%%%%%%%%%%%%%%%%%%%

%%%%%%%%%%%%%%%%%%%%%%%%%%%%%%%%%%%%%%%%%%%%%%%%%
\begin{figure*}
\centering
\includegraphics[width=0.8\textwidth]{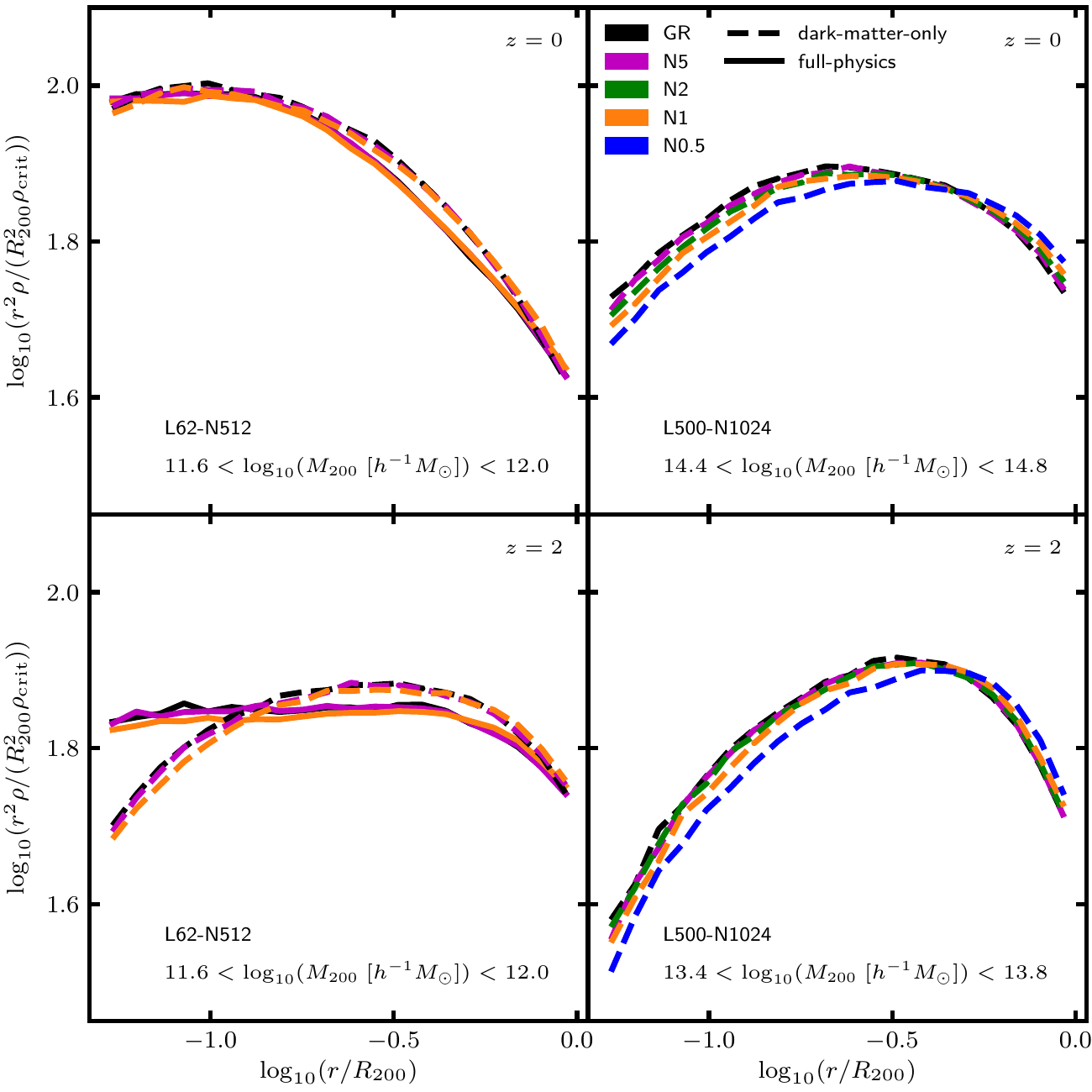}
\caption{[{\it Colour Online}] Median density profiles of haloes from the \textsc{arepo} simulations L62 (\textit{left column}) and L500 (\textit{right column}) at redshifts $0$ (\textit{top row}) and $2$ (\textit{bottom row}). Data from both the full-physics (\textit{solid lines}) and dark-matter-only (\textit{dashed lines}) counterparts of L62 are shown. The L500 simulation includes runs for GR (\textit{black}) and the nDGP models N5 (\textit{magenta}), N2 (\textit{green}), N1 (\textit{orange}) and N0.5 (\textit{blue}), while the L62 simulation includes GR, N5 and N1 only. The mass bins used to measure the median density are annotated.}
\label{fig:density}
\end{figure*}
%%%%%%%%%%%%%%%%%%%%%%%%%%%%%%%%%%%%%%%%%%%%%%%%%%

%%%%%%%%%%%%%%%%%%%%%%%%%%%%%%%%%%%%%%%%%%%%%%%%%
\begin{figure*}
\centering
\includegraphics[width=1.0\textwidth]{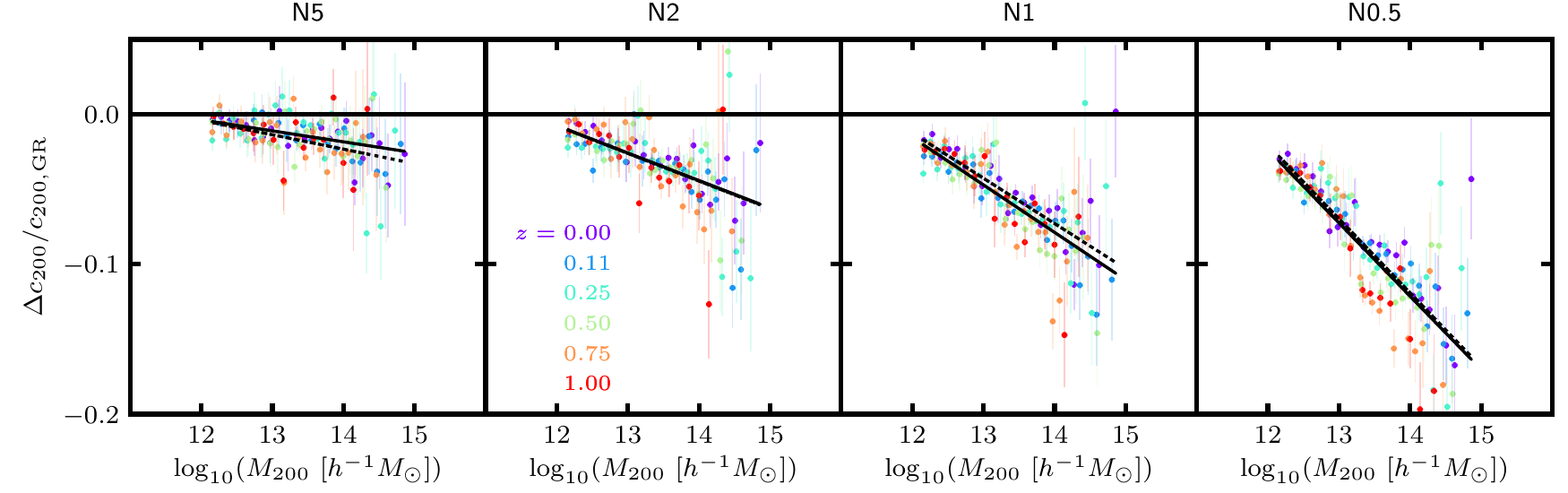}
\caption{[{\it Colour Online}] Relative difference between the median halo concentration in nDGP with respect to GR, as a function of the mean logarithm of the halo mass. Binned data are shown for all snapshots with redshift $z\leq1$, where the redshift is represented by colour. The data is generated using the dark-matter-only simulations L62, L200 and L500, the specifications of which are given in Table \ref{table:simulations}. These have been run for GR and the nDGP models N5, N2, N1 and N0.5 (shown from left to right). The error bars indicate the $1\sigma$ uncertainties. The solid lines represent the best-fit linear relations for each panel, while the dashed lines show the predictions from our general model, which is given by Eq.~(\ref{eq:c_model}). For all models, and across a halo mass range of four orders of magnitude, the fitting function gives a percent-level agreement with the simulation measurement of the concentration decrement at $0\leq z\leq1$.}
\label{fig:linear_fit}
\end{figure*}
%%%%%%%%%%%%%%%%%%%%%%%%%%%%%%%%%%%%%%%%%%%%%%%%%%

%%%%%%%%%%%%%%%%%%%%%%%%%%%%%%%%%%%%%%%%%%%%%%%%%
\begin{figure*}
\centering
\includegraphics[width=1.0\textwidth]{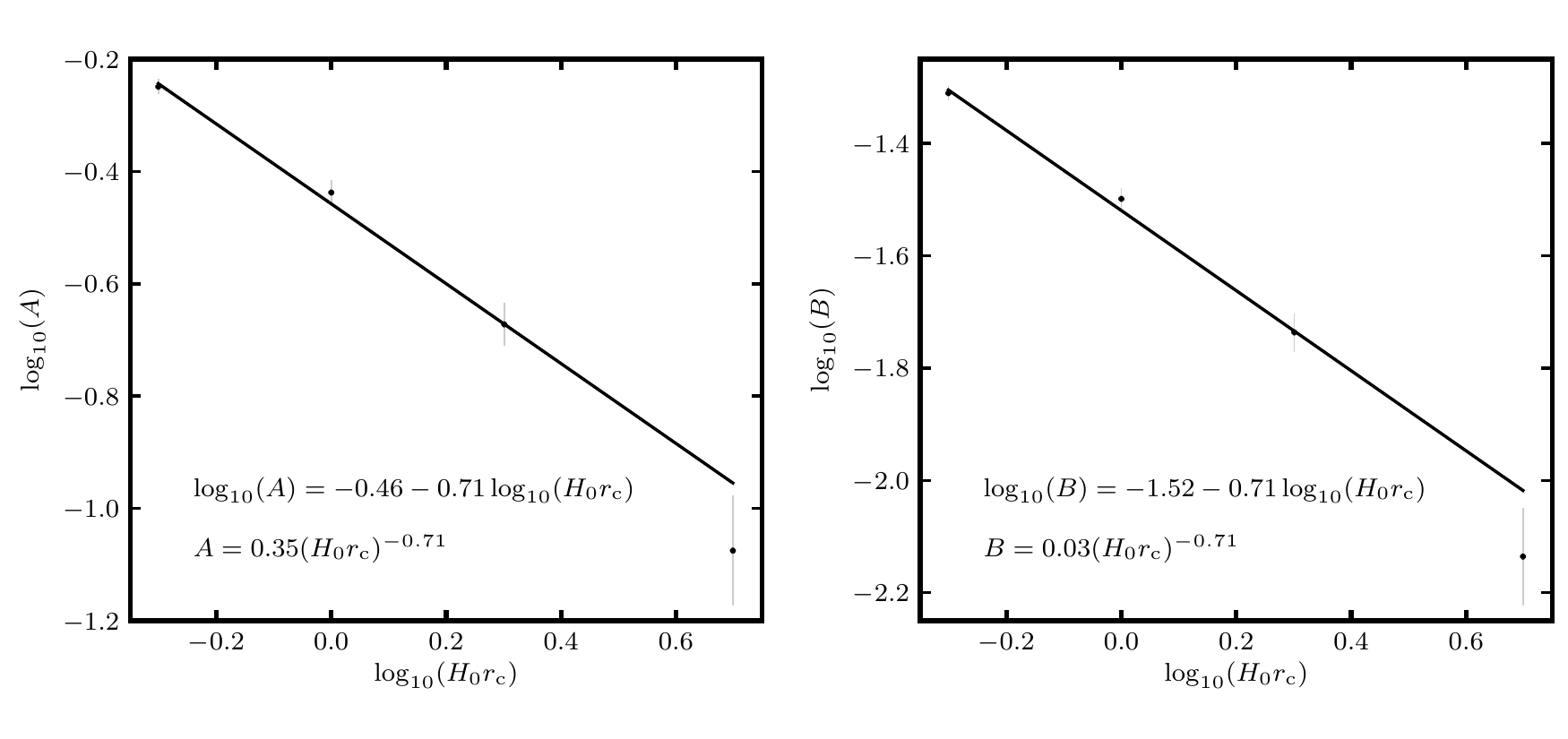}
\caption{[{\it Colour Online}] Best-fit values of the parameters $A$ and $B$ of Eq.~(\ref{eq:linear_fit}) as a function of the logarithm of $H_0r_{\rm c}$, where $r_{\rm c}$ is the cross-over scale of nDGP gravity. The best-fit values of the data points have been computing by fitting Eq.~(\ref{eq:linear_fit}) to the data shown in the four panels of Fig.~\ref{fig:linear_fit}, which correspond to the models N5, N2, N1 and N0.5. The error bars represent the $1\sigma$ uncertainties in the data, obtained from the weighted least squares fits. The solid lines show best-fit power-law fits of the four data points, Eq.~(\ref{eq:power_laws}), which are annotated.}
\label{fig:power_law}
\end{figure*}
%%%%%%%%%%%%%%%%%%%%%%%%%%%%%%%%%%%%%%%%%%%%%%%%%%

In Sec.~\ref{sec:results:c_DMO}, we discuss the concentration results from our DMO simulations (dashed lines in Figs.~\ref{fig:c_M}-\ref{fig:density}). Then, in Sec.~\ref{sec:results:c_FP}, we summarise the results from full-physics simulations (solid lines in Figs.~\ref{fig:c_M}-\ref{fig:density}), including the effect of baryons on the model differences. Finally, in Sec.~\ref{sec:results:concentration_model}, we present a general model for the concentration-mass relation in nDGP.

\subsubsection{Dark-matter-only concentration}
\label{sec:results:c_DMO}

In order to study the concentration over a wide and continuous halo mass range, we have combined the data from our DMO simulations into a single catalogue. In order to avoid resolution issues with the concentration measurement, we exclude haloes which have fewer than 2000 particles (within the radius $R_{200}$) and we leave out L1000 due to its low mass resolution. The resulting catalogue consists of haloes spanning masses $3.04\times10^{11}h^{-1}M_{\odot}\lesssim M_{200}\lesssim10^{15}h^{-1}M_{\odot}$. We note that, because L62 has not been run for N2 and N0.5, the data for these models only extends down to mass $1.278\times10^{12}h^{-1}M_{\odot}$ ($\equiv2000$ particles from L200). Throughout this section, we will only refer to the results from this combined catalogue; however, in Appendix \ref{sec:appendix:consistency}, we also compare the concentration predictions from each of our DMO simulations, including L1000.

The top row of Fig.~\ref{fig:c_M} shows the median concentration as a function of mass for redshifts $0$, $1$ and $2$ (from left to right). The median has been computed using mass bins containing a minimum of 100 GR haloes each: the bins all have equal width in logarithmic mass apart from the highest-mass bin, which is wide enough to enclose the 100 highest-mass haloes. The same set of bins is used for each gravity model. As expected from literature \citep[e.g.,][]{Duffy:2008pz}, the median concentration appears to follow a descending power-law relation with the mass. This behaviour arises due to the hierarchical nature of structure formation: higher-mass haloes form at later times when the background density is lower. Therefore, the concentration of these haloes is also typically lower.

The bottom row of Fig.~\ref{fig:c_M} shows the relative difference between the nDGP and GR median concentrations. The shaded region shows the $1\sigma$ error. To calculate this, the standard error of the mean (equal to the standard deviation divided by the square root of the halo count) is computed for each mass bin for GR and nDGP, and then combined in quadrature. We note that, although the nDGP and GR simulations are started from the same initial conditions, the differing gravitational forces affect the trajectories of the simulation particles, which end up at different positions with different velocities, essentially losing much of the memory of their initial states. Therefore, the concentration measurements of each model can be treated as independent, so that the errors may be combined as described. Our results show that the nDGP fifth force causes the concentration to be reduced, since particles experience the fifth force and hence have enhanced velocities before they fall into haloes, so that after entering the haloes their higher kinetic energy makes it harder for them to settle towards the central regions. The effect is greater for models which have a stronger fifth force, so the concentration suppression is highest in N0.5 ($\sim${$10\%$} on average) and lowest in N5 (at percent level). At $z=0$ and $z=1$, the suppression is greater at higher mass. This appears to be the case for N0.5 at $z=2$ as well, but not for weaker models, where the suppression appears to have a much weaker dependence on the halo mass.

To complement these results, we show the median concentration, computed within three mass bins, as a function of redshift in Fig.~\ref{fig:c_z}. The lower-mass bin, $10^{11.6}h^{-1}M_{\odot}<M_{200}<10^{12}h^{-1}M_{\odot}$, corresponds to galaxy-sized haloes: here, we use haloes from L62, for which we again note that only the GR, N5 and N1 models are available. For the middle-mass bin, $10^{12.8}h^{-1}M_{\odot}<M_{200}<10^{13.2}h^{-1}M_{\odot}$, we use haloes from L200. For both of these bins, the nDGP suppression of the concentrations appears to be approximately constant over the redshift range $0\leq z\leq3$, ranging from a couple of percent at most in N5 to about $7\%$ in N0.5.

The higher-mass bin,  $10^{14}h^{-1}M_{\odot}<M_{200}<10^{14.4}h^{-1}M_{\odot}$, shown in Fig.~\ref{fig:c_z} corresponds to cluster-sized objects; for this, we use haloes from L500. Because clusters typically form at later times, this bin consists of fewer than 100 haloes for redshifts $z\gtrsim1.25$, and we therefore exclude these redshifts from the figure. The suppression of the concentration in nDGP is greater for this bin than for the lower-mass bins, reaching $\sim15\%$ in N0.5. This is consistent with the results of Fig.~\ref{fig:c_M}. As for the other bins, the suppression does not appear to evolve with redshift in N5, N2 and N1. However, for N0.5, the suppression is slightly greater at $z=1$ ($\sim15\%$), than at $z=0$ ($\sim12\%$). We note that the error is also greater at high redshift due to the reduced number of objects, so these results alone do not provide compelling evidence of a redshift evolution of the concentration suppression.

To help make sense of these results, in Fig.~\ref{fig:density} we show the median density profiles of haloes from a few mass bins at redshifts 0 and 2. These have been computed by measuring the median density, in radial bins spanning $0.05R_{200}$ to $R_{200}$, using the binned haloes. The density has been scaled by $r^2$ so that the profiles peak at the scale radius, $R_{\rm s}$. This means that the concentration, $c_{200}=R_{200}/R_{\rm s}$, can effectively be read off from the peak radius: a higher (lower) peak radius corresponds to a lower (higher) concentration. In the left column of Fig.~\ref{fig:density}, we show the median profile for haloes from L62 in the mass bin $10^{11.6}h^{-1}M_{\odot}<M_{200}<10^{12}h^{-1}M_{\odot}$. In the right column, we use haloes from L500 within mass bins $10^{14.4}h^{-1}M_{\odot}<M_{200}<10^{14.8}h^{-1}M_{\odot}$ and $10^{13.4}h^{-1}M_{\odot}<M_{200}<10^{13.8}h^{-1}M_{\odot}$ at redshifts 0 and 2, respectively. We use a lower mass for the $z=2$ profile due to the limited number of haloes at higher masses.

For the higher mass bins -- where we have seen that there is a greater suppression of the concentration in nDGP models -- a clear trend is present: at the outer (inner) regions of haloes, the density is greater (lower) in nDGP than in GR. As mentioned above, this is related to the nature of the Vainshtein screening in nDGP, which suppresses the fifth force on small scales or distances. This means that the fifth force is stronger at large scales, which correspond to the outer regions of these haloes and regions further away from the halo-formation sites. This causes orbiting dark matter particles to undergo an enhanced gravitational acceleration at these regions and have higher kinetic energy, which prevents them from relaxing and settling into lower-radius orbits where the fifth force is suppressed. This causes $r^2\rho(r)$ to peak at a higher radius in the nDGP models than in GR, resulting in a suppressed concentration. The effect is greatest in N0.5.

For the lower mass bins, we have seen in Figs.~\ref{fig:c_M} and \ref{fig:c_z} that the effect of the fifth force is not as strong. This is consistent with the low-mass density profiles in Fig.~\ref{fig:density}, where the nDGP profiles are closer to GR. However, the density is still slightly reduced at the inner regions and increased at the outer regions, and so the concentration is still suppressed. The reason that the effect is not as strong at low mass is again due to the nature of the Vainshtein screening: lower-mass haloes have a smaller spatial extent, therefore the small-scale suppression of the fifth force is more substantial throughout the range $r<R_{200}$. In addition, smaller haloes generally form at higher redshifts, so that the particles inside them have spent less time outside the haloes and are therefore less affected by the fifth force; this is because, once these particles enter haloes, the fifth force is strongly suppressed.

\subsubsection{Full-physics concentration}
\label{sec:results:c_FP}

In Figs.~\ref{fig:c_M}-\ref{fig:density}, we have also included data from our full-physics simulations, which are represented with solid lines. Because these data are only available for the $62h^{-1}{\rm Mpc}$ box, the data only extends to low-mass galaxy clusters (although, we note that the mean logarithmic mass of the rightmost bin shown in Fig.~\ref{fig:c_M} is only slightly above $10^{13}h^{-1}M_{\odot}$). Nevertheless, by comparing this to the data from the combined DMO data, we can get an idea of how the results differ when gas and processes such as star formation and feedbacks are included.

From the solid lines in Figs.~\ref{fig:c_M} and \ref{fig:c_z}, we see that the full-physics concentration is typically greater at lower masses and reduced at higher masses. The full-physics simulations include a gaseous component which, unlike dark matter, is affected by turbulence. This causes the gas particles to slow down and settle at the inner regions of haloes. Also, at the centre of a halo, we are likely to see stellar particles concentrate. This means that the total halo density is enhanced in the inner regions, which is consistent with the stacked density profiles of the full-physics simulations in Fig.~\ref{fig:density}. According to these results, the rescaled density profile becomes approximately flat at the inner regions, corresponding to a $\rho(r)\propto r^{-2}$ power-law. This clearly deviates from the NFW profile, which follows an $r^{-1}$ power law in these regions. Because the concentration is a parameter of the NFW profile, we still have to fit Eq.~(\ref{eq:nfw_log}) in order to measure this. Doing so produces a value that is either higher or lower than for DMO haloes with the same mass.

Despite the difference in the absolute concentrations, the suppression of the concentration in nDGP appears to have a similar magnitude in the full-physics and DMO simulations, according to the bottom-left panel of Fig.~\ref{fig:c_z}, for galaxy-sized haloes in the redshift range $0\leq z\leq3$. In Fig.~\ref{fig:c_M}, the dashed and solid lines in the lower panels also appear to have a similar magnitude; however, we are unable to rigorously test this for masses $M_{200}\gtrsim10^{13}h^{-1}M_{\odot}$, which would require full-physics simulations of nDGP that have a much larger box size. Such simulations are highly expensive, and are therefore left for future work.

\subsubsection{Modelling the concentration in nDGP}
\label{sec:results:concentration_model}

From Figs.~\ref{fig:c_M} and \ref{fig:c_z}, it appears that the suppression of the DMO halo concentration in nDGP grows with mass and is approximately constant as a function of redshift. In Fig.~\ref{fig:linear_fit}, we show the binned relative difference data from our combined DMO simulation data for all snapshots at $z\leq1$. The data appears to follow a linear trend as a function of the mass, therefore we can model this using:
\begin{equation}
    \Delta c/c_{\rm GR} = A - B\log_{10}(M_{200}M_{\odot}^{-1}h),
    \label{eq:linear_fit}
\end{equation}
where $A$ and $B$ are parameters representing the amplitude and slope of the relation, respectively. This does not include any dependence on redshift. For the N0.5 data, there is a clear $z$-dependence, with low-$z$ (blue) data having a smaller suppression than high-$z$ (red) data; however, the suppression in different snapshots is still quite close, and there does not appear to be any $z$-evolution for the other, more realistic, models of nDGP.

The solid lines in Fig.~\ref{fig:linear_fit} are the best-fit relations for each model. These are created by using weighted least squares to fit Eq.~(\ref{eq:linear_fit}) to the data points, where points with large (small) error bars are given smaller (larger) weighting. In Fig.~\ref{fig:power_law}, we show the best-fit values of $A$ and $B$ as a function of the $H_0r_{\rm c}$ parameter which characterises the nDGP models. Both $A$ and $B$ appear to be well-described by a power-law relation. Using weighted least squares to fit the four data points, we obtain the following best-fit relations:
\begin{equation}
    \begin{split}
    &A = (0.35\pm0.01)(H_0r_{\rm c})^{-0.71\pm0.05};\\
    &B = (0.0302\pm0.0008)(H_0r_{\rm c})^{-0.71\pm0.05}.
    \end{split}
    \label{eq:power_laws}
\end{equation}
Interestingly, the relations both have power-law slope $-0.71\pm0.05$. They can therefore be combined with Eq.~(\ref{eq:linear_fit}) to form the following simple relation:
\begin{equation}
    \begin{split}
    \frac{\Delta c}{c_{\rm GR}} = &[(0.35\pm0.01) - (0.0302\pm0.0008)\log_{10}(M_{200}M_{\odot}^{-1}h)]\\
    &\times(H_0r_{\rm c})^{-0.71\pm0.05}.
    \end{split}
    \label{eq:c_model}
\end{equation}
This %model 
can be used to predict the suppression of the concentration in nDGP, as a function of the halo mass $M_{200}$ and model parameter $H_0r_{\rm c}$. The dashed lines in Fig.~\ref{fig:linear_fit} show the model predictions for our four nDGP models. The agreement with the data is generally very good for the full mass range, $10^{12}h^{-1}M_{\odot}\lesssim M_{200} \lesssim 10^{15}h^{-1}M_{\odot}$, of our simulation data. The agreement is particularly good for weaker models, where it appears to match $z=0$ and $z=1$ data equally well. For N0.5, which is our strongest model, our relation appears to slightly underestimate the concentration suppression for high-redshift data; however, the overall level of agreement is still very good, considering that the model is able to give reasonable predictions for such a wide range of nDGP models and masses.

%%%%%%%%%%%%%%%%%%%%%%%%%%%%%%%%%%%%%%%%%%%%%%%%%
\begin{figure*}
\centering
\includegraphics[width=1.0\textwidth]{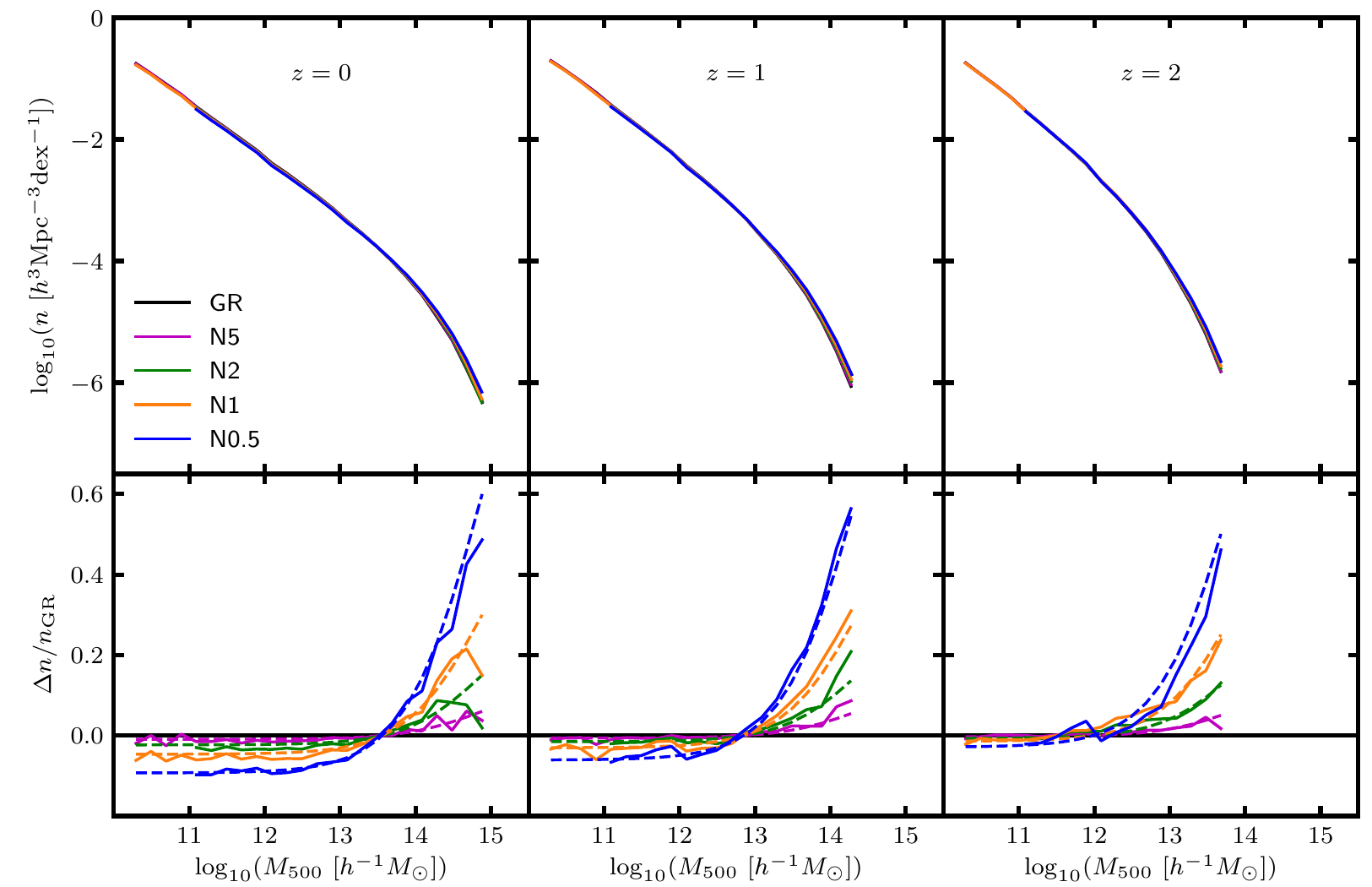}
\caption{[{\it Colour Online}] Halo mass function (\textit{top row}) and its relative difference in nDGP with respect to GR (\textit{bottom row}), as a function of the mean logarithm of the halo mass at redshifts $0$, $1$ and $2$. The data is generated using our \textsc{arepo} dark-matter-only simulations, the specifications of which are given in Table \ref{table:simulations}. These have been run for GR (\textit{black}) and the nDGP models N5 (\textit{magenta}), N2 (\textit{green}), N1 (\textit{orange}) and N0.5 (\textit{blue}). The dashed lines show the predictions from our general fitting model, which is given by Eqs.~(\ref{eq:hmf}, \ref{eq:hmf_params}).}
\label{fig:hmf_combined}
\end{figure*}
%%%%%%%%%%%%%%%%%%%%%%%%%%%%%%%%%%%%%%%%%%%%%%%%%%

\subsection{Halo mass function}
\label{sec:results:hmf}

The HMF does not have a strict mass resolution requirement like the concentration, therefore we use all haloes which have at least 100 particles within the radius $R_{500}$. We again combine the halo data from our DMO simulations, and the relaxed resolution requirement means that we can now also include L1000 haloes. The HMF is computed using mass bins with equal logarithmic width 0.2. The halo count in each bin is divided by the total volume from all contributing simulations: for example, the volume is $62^3h^{-3}{\rm Mpc}^3$ for the lowest-mass bins where only the L62 box has sufficient resolution, and $(62^3+200^3+500^3+1000^3)h^{-3}{\rm Mpc}^3$ for the highest-mass bins where all simulations have sufficient resolution. In Appendix \ref{sec:appendix:consistency}, we also assess the consistency of our DMO simulations by comparing the HMF predictions at different resolutions.

The binned HMF is shown in Fig.~\ref{fig:hmf_combined} for redshifts 0, 1 and 2, where only mass bins containing at least 100 haloes are displayed. We note that, because our highest-resolution simulation L62 has been run for N5 and N1 only, the data for these models extends to lower masses than the other models. The relative difference between the nDGP and GR results is shown in the lower panels. For all three redshifts, the HMF is significantly enhanced in nDGP relative to GR at high mass: for N0.5, the HMF is enhanced by up to $60\%$, while for N5 the enhancement is less than $10\%$. On the other hand, the HMF is suppressed at lower masses in nDGP, by up to $\sim10\%$ in N0.5 and a couple of percent in N5. The threshold mass above which the HMF is enhanced and below which it is suppressed is higher at lower redshifts, with values $\sim10^{13.5}h^{-1}M_{\odot}$ at $z=0$ and $\sim10^{12}h^{-1}M_{\odot}$ at $z=2$. The low-mass suppression of the HMF also decreases with redshift.

These results can again be explained by the behaviour of the fifth force, which enhances the overall strength of gravity on large scales, accelerating the formation of high-mass haloes so that there is a greater abundance of these objects in nDGP compared to GR at a given time. On the other hand, the abundance of low-mass haloes, which undergo an increased number mergers, is reduced. The mass threshold between HMF enhancement and HMF suppression is reduced at higher redshifts, which is likely simply because the masses of a given population of haloes are lower at earlier times. 

Structure formation is sped up by a greater extent in models which feature a stronger fifth force, so the effects described above are greater for N0.5 than for weaker models. The enhancement of the HMF is greatest at the high-mass end. Therefore, by using observations of high-mass galaxy clusters from ongoing and upcoming galaxy surveys \citep[e.g.,][]{desi,euclid,lsst}, it will be possible to make powerful constraints of nDGP. However, any tests of the nDGP model of this kind may be affected by the cluster observable-mass scaling relations discussed earlier, and we will investigate the implications of this in a follow-up work.

From the lower panels of Fig.~\ref{fig:hmf_combined}, it appears that, for any model, i.e., for a given choice of $H_0r_{\rm c}$, the HMF enhancement has a constant shape, but shifts downwards and towards larger $M_{500}$ as one goes to lower redshifts. Therefore, it can be well-described by the following model:
\begin{equation}
    \frac{\Delta n}{n_{\rm GR}} = A(H_0r_{\rm c})\left[\tanh\left(\log_{10}(M_{500}M_{\odot}^{-1}h) - B(z)\right) + C(z)\right].
    \label{eq:hmf}
\end{equation}
We use a portion of a $\tanh$ function to represent the mass-dependent shape, which is level at low mass and rises steeply at high mass. We also include the following parameters: $A(H_0r_{\rm c})$ controls the amplitude, which depends on the model parameter $H_0r_{\rm c}$; $B(z)$ represents the $z$-dependent shift along the mass axis; and $C(z)$ represents the $z$-dependent shift along the $\Delta n/n_{\rm GR}$ axis. By adopting simple linear models for each of these parameters, and by combining the data from all simulation snapshots in the range $0\leq z\leq2$, we have used unweighted least squares to obtain the following best-fit results:
\begin{equation}
    \begin{split}
    &A(H_0r_{\rm c}) = (0.342\pm0.014){\left(H_0r_{\rm c}\right)^{-1}},\\
    &B(z) = (14.87\pm0.03) - (0.481\pm0.010)z,\\
    &C(z) = (0.864\pm0.008) + (0.047\pm0.005)z.
    \end{split}
    \label{eq:hmf_params}
\end{equation}
The predictions of this calibrated model are indicated by the dashed lines in Fig.~\ref{fig:hmf_combined}. The agreement with the simulation data is excellent for all models for the mass ranges shown, which span $4$--$5$ decades depending on redshift. At $z=0$ and $z=1$, apart from the highest mass bin where data is noisy, the agreement between the fitting function and simulation measurements is within $\sim3\%$; at $z=2$, the agreement is within $\sim3\%$ for all but the strongest model ($H_0r_{\rm c}=0.5$) where we still have a $\lesssim5\%$ accuracy. In the limit $H_0r_{\rm c}\rightarrow\infty$, where nDGP becomes GR, our model predicts a relative difference of zero as expected. However, we note that our model will predict a constant relative difference if extrapolated to higher masses. This behaviour may not be physically accurate, but the high halo masses are beyond the dynamical range of our simulations and so we cannot test this reliably. Therefore, the model in Eqs.~(\ref{eq:hmf}, \ref{eq:hmf_params}) should only be used for the mass range $10^{11}h^{-1}M_{\odot}\lesssim M_{500}\lesssim M_{\rm max}(z)$, where $M_{\rm max}(z)$ is the maximum mass used to calibrate the above model at a given redshift. The latter can be estimated using the relation:
\begin{equation}
    \log_{10}\left(M_{\rm max}M_{\odot}^{-1}h\right) = 14.81 - 0.54z,
\end{equation}
which we have calibrated using snapshots in the range $0\leq z\leq2$. 

In this section, we have focused on the mass definition $M_{500}$, which is commonly used in cluster number counts studies \citep[e.g.,][]{Planck_SZ_cluster}. For completeness, we also present, in Appendix \ref{sec:appendix:hmf}, results and modelling for mass definition $M_{200}$.

\section{Summary, Discussion and Conclusions}
\label{sec:conclusions}

Since the first detection of the accelerated expansion of the cosmos, a wide variety of MG theories have been proposed which can give rise to this late-time phenomenon. These models often feature a `fifth force', which can alter the formation of structure on cosmological scales. This can, for example, affect the number density of galaxy clusters, which then offers a powerful probe for constraining these theories. It is an exciting time for cluster cosmology, with a wealth of high-quality data being made available from ongoing and upcoming surveys \citep[e.g.,][]{desi,erosita,lsst}.

This paper is part of a series of works dedicated to developing a framework for making the best-use of these cluster observations by obtaining robust and unbiased constraints of MG theories. So far, we have focused on HS $f(R)$ gravity, in which the strength of gravity is enhanced in sufficiently low-density regimes. In this paper, we have extended this framework to the popular nDGP model, in which a fifth force is able to act over sufficiently large scales.

Using the first cosmological simulations that simultaneously incorporate full baryonic physics and the nDGP model, we have studied the observable-mass scaling relations for three mass proxies (see Sec.~\ref{sec:results:scaling_relations}). For groups and clusters in the mass range $M_{500}\lesssim10^{14.5}M_{\odot}$, our results show that for the N1 model, the $\bar{T}_{\rm gas}(M)$ relation is enhanced by about 5\% with respect to GR, while the $Y_{\rm SZ}(M)$ and $Y_{\rm X}(M)$ relations are both enhanced by 10\%-15\% at low masses but more closely match the GR relations at high masses. For N5, which is much weaker than N1, the $\bar{T}_{\rm gas}(M)$ relation closely resembles the GR relation, while the $Y_{\rm SZ}(M)$ and $Y_{\rm X}(M)$ relations are enhanced by up to 5\% at low mass and suppressed by up to 5\% at high mass. These deviations from GR could be related to the effect of the fifth force on gas velocities during cluster formation, and they also hint at an interplay between the fifth force and stellar and black hole feedback.

Using a suite of DMO $N$-body simulations, which cover a wide range of resolutions and box sizes, we have found that, in nDGP, the concentration is typically suppressed relative to GR, varying from a few percent in N5 to up to $\sim15\%$ in N0.5 (see Sec.~\ref{sec:results:concentration}). Using stacked density profiles at different mass bins, we have shown that this behaviour is caused by a reduced (increased) density at the inner (outer) halo regions. Including full baryonic physics significantly affects the concentration-mass relation; however, our results show that, for masses $M_{200}\lesssim10^{13}h^{-1}M_{\odot}$, the model differences between nDGP and GR still have a similar magnitude compared to the DMO simulations.

By combining the data from our $z\leq1$ simulation snapshots, we have calibrated a general model, given by Eq.~(\ref{eq:c_model}), which is able to accurately predict the suppression of the halo concentration with respect to the GR results as a function of the halo mass and the $H_0r_{\rm c}$ parameter of nDGP over ranges $10^{12}h^{-1}M_{\odot}\lesssim M_{200}\lesssim 10^{15}h^{-1}M_{\odot}$ and 0.5-5, respectively. This model can be included in our MCMC pipeline for converting between mass definitions in case, for example, the theoretical predictions and observables are defined with respect to different spherical overdensities. Our model can also be used, along with the HMF, to predict the nonlinear matter power spectrum, which can also be used to constrain gravity.

We have also used our DMO simulations to study the HMF over the mass range $1.52\times10^{10}h^{-1}M_{\odot}\leq M_{500}\lesssim 10^{15}h^{-1}M_{\odot}$ at redshifts 0, 1 and 2 (see Sec.~\ref{sec:results:hmf}). Our results (Fig.~\ref{fig:hmf_combined}), indicate that the nDGP HMF is enhanced at high masses (by up to $\sim60\%$ in N0.5) and suppressed at low masses (by $\sim10\%$ in N0.5) compared to GR. These results indicate the potential constraining power from using the observed mass function to probe the $H_0r_{\rm c}$ parameter of nDGP. By combining the data from our $z\leq2$ snapshots, we have calibrated a general model, given by Eq.~(\ref{eq:hmf}), which can accurately reproduce the HMF enhancement as a function of the halo mass, redshift and $H_0r_{\rm c}$ parameter. This model can be used for theoretical predictions of the nDGP HMF (using a parameter-dependent GR calibration) in our MCMC pipeline.

In \citet{Mitchell:2020aep}, we showed that a model for the $f(R)$ dynamical mass enhancement can be used to predict observable-mass scaling relations in $f(R)$ gravity using their GR counterparts. Such a model in nDGP could similarly be useful to help understand the enhancements of the temperature and SZ and X-ray $Y$-parameters observed in this work. This is left to a future study. For now, though, we note that the scaling relations in nDGP still appear to follow power-law relations as a function of the mass: the $\bar{T}_{\rm gas}(M)$ relation in N1 can be related to the GR relation by a simple rescaling of the amplitude, whereas the $Y_{\rm SZ}(M)$ and $Y_{\rm X}(M)$ relations appear to have shallower slopes in N5 and N1 than in GR. Therefore, in our future MCMC pipeline for obtaining constraints of nDGP, we can still assume the GR power-law form of the scaling relations by allowing the parameters controlling the amplitude and slope to vary along with the cosmological and nDGP parameters \citep[e.g.,][]{deHaan:2016qvy,Bocquet:2018ukq}. 

Although our simulations have only been run for a single choice of cosmological parameters, we expect that our models for the enhancements of the halo concentration and HMF will have a reasonable accuracy for other (not too exotic) parameter values. The gravitational force enhancement in nDGP, given by $\left[1 + 1/(3\beta)\right]$, has only a weak dependence on $\Omega_{\rm M}$: for the N1 model ($\Omega_{\rm rc}=0.25$), the force enhancement varies within a very small range (roughly $12.1\%-12.6\%$) for $\Omega_{\rm M} \in [0.25,0.35]$ at the present day, and the range of variation is even smaller at higher redshifts. Therefore, for now we assume that the effects of the cosmological parameters on the concentration and HMF are approximately cancelled out in the ratios $\Delta c/c_{\rm GR}$ and $\Delta n/n_{\rm GR}$. However, we will revisit this in a future work, using a large number of nDGP simulations that are currently being run for different combinations of cosmological parameters, before these models are used in tests of gravity using observational data.

Finally, we note that, because the \textsc{shybone} simulations have a small box size ($62h^{-1}{\rm Mpc}$), it is difficult to robustly model the observable-mass scaling relations for cluster-sized objects ($M_{500}\gtrsim10^{14}M_{\odot}$). It would therefore be useful to revisit this study using full-physics nDGP simulations with a larger box. We have been fine-tuning a new baryonic model which can allow TNG-like simulations to be run at a much lower resolution, making it possible to run large simulations with reduced computational cost. We will present this model in an upcoming work, in which we will also revisit our $f(R)$ scaling relation results using much larger simulations.

\section*{Acknowledgements}

MAM is supported by a PhD Studentship with the Durham Centre for Doctoral Training in Data Intensive Science, funded by the UK Science and Technology Facilities Council (STFC, ST/P006744/1) and Durham University. CA and BL are supported by the European Research Council via grant ERC-StG-716532-PUNCA. BL is additionally supported by STFC Consolidated Grants ST/T000244/1 and ST/P000541/1. CH-A is supported by the Excellence Cluster ORIGINS which is funded by the Deutsche Forschungsgemeinschaft (DFG, German Research Foundation) under Germany's Excellence Strategy - EXC-2094 - 390783311. This work used the DiRAC@Durham facility managed by the Institute for Computational Cosmology on behalf of the STFC DiRAC HPC Facility (\url{www.dirac.ac.uk}). The equipment was funded by BEIS capital funding via STFC capital grants ST/K00042X/1, ST/P002293/1, ST/R002371/1 and ST/S002502/1, Durham University and STFC operations grant ST/R000832/1. DiRAC is part of the National e-Infrastructure.

\section*{Data availability}

The simulation data and results of this paper may be available upon request.

%%%%%%%%%%%%%%%%%%%%%%%%%%%%%%%%%%%%%%%%%%%%%%%%%%

%%%%%%%%%%%%%%%%%%%% REFERENCES %%%%%%%%%%%%%%%%%%

% The best way to enter references is to use BibTeX:

\bibliographystyle{mnras}
\bibliography{references} % if your bibtex file is called example.bib

% Alternatively you could enter them by hand, like this:
% This method is tedious and prone to error if you have lots of references

%%%%%%%%%%%%%%%%%%%%%%%%%%%%%%%%%%%%%%%%%%%%%%%%%%

%%%%%%%%%%%%%%%%% APPENDICES %%%%%%%%%%%%%%%%%%%%%
\appendix

\section{Simulation consistency}
\label{sec:appendix:consistency}

%%%%%%%%%%%%%%%%%%%%%%%%%%%%%%%%%%%%%%%%%%%%%%%%%
\begin{figure*}
\centering
\includegraphics[width=0.85\textwidth]{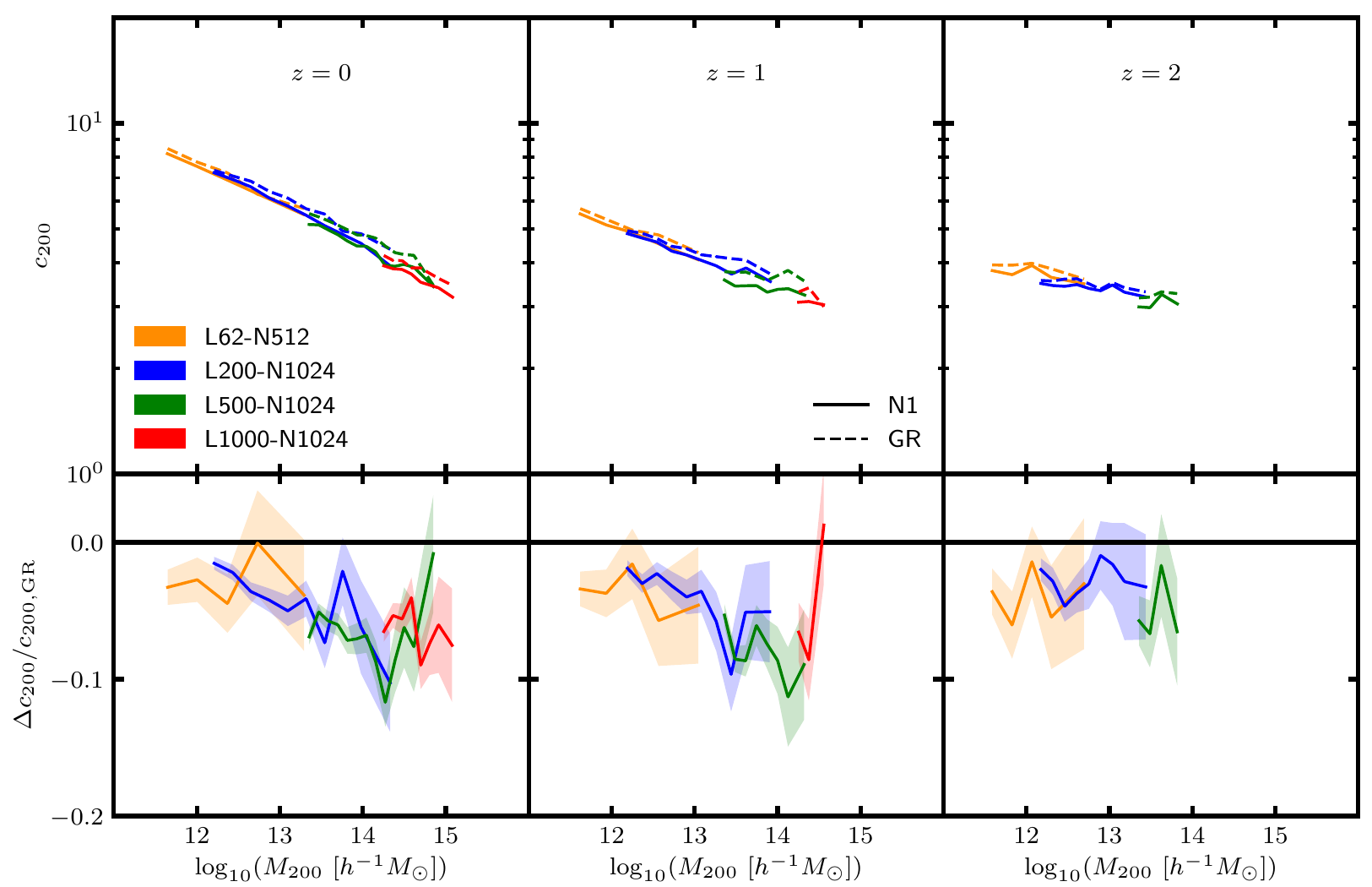}
\caption{[{\it Colour Online}] Median halo concentration (\textit{top row}) and relative difference with respect to GR (\textit{bottom row}) as a function of the mean logarithm of the halo mass at redshifts $0$, $1$ and $2$. The data is generated using the dark-matter-only simulations L62 (\textit{orange}), L200 (\textit{blue}), L500 (\textit{green}) and L1000 (\textit{red}), the specifications of which are given in Table \ref{table:simulations}. Data is shown for GR (\textit{dashed lines}) and the nDGP model N1 (\textit{solid lines}). The shaded regions in the lower panels show the $1\sigma$ uncertainty in the relative difference.}
\label{fig:c_consistency}
\end{figure*}
%%%%%%%%%%%%%%%%%%%%%%%%%%%%%%%%%%%%%%%%%%%%%%%%%%

%%%%%%%%%%%%%%%%%%%%%%%%%%%%%%%%%%%%%%%%%%%%%%%%%
\begin{figure*}
\centering
\includegraphics[width=0.85\textwidth]{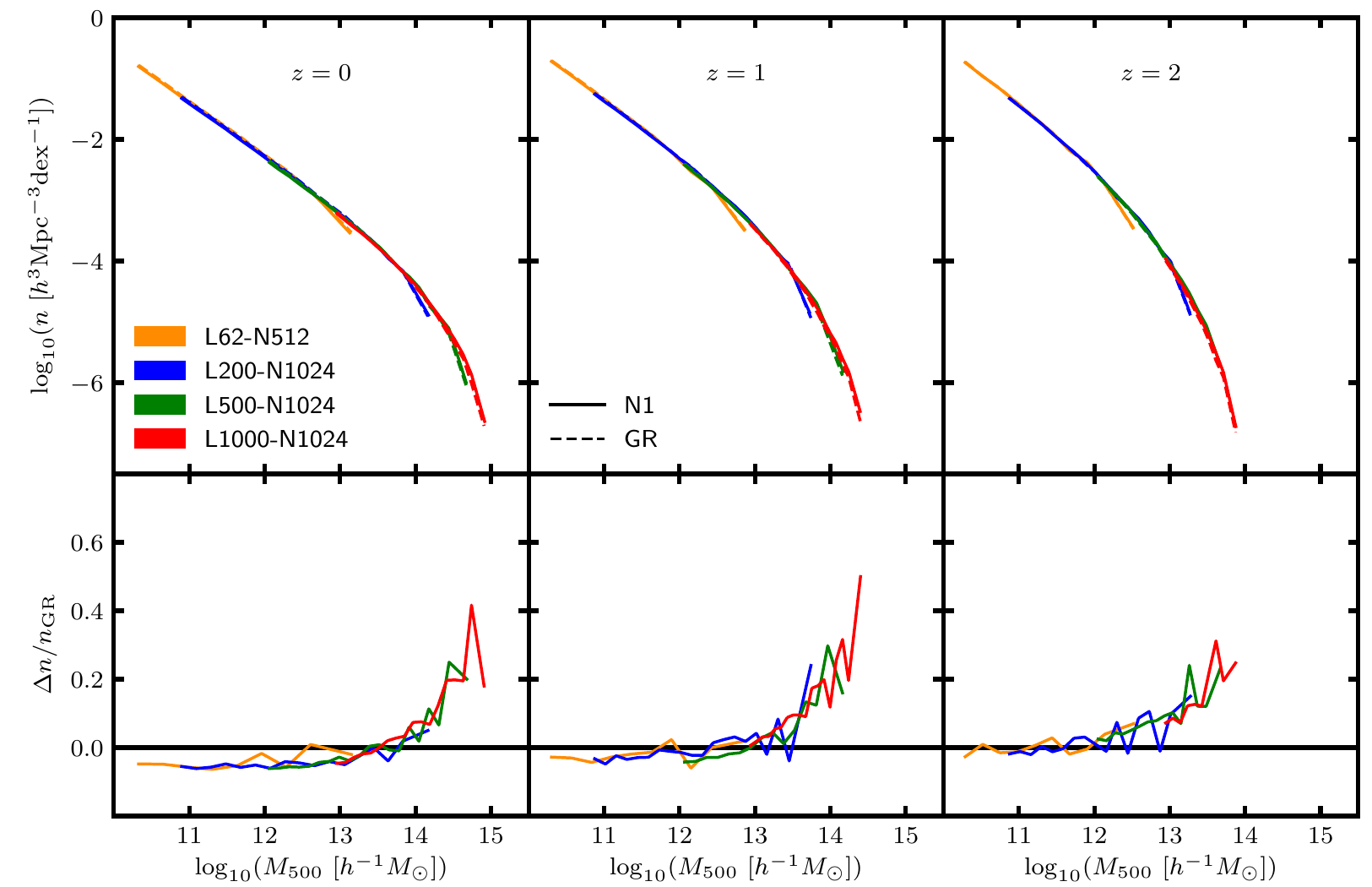}
\caption{[{\it Colour Online}] Halo mass function (\textit{top row}) and its relative difference in nDGP with respect to GR (\textit{bottom row}), as a function of the mean logarithm of the halo mass at redshifts $0$, $1$ and $2$. The data is generated using the dark-matter-only simulations L62 (\textit{orange}), L200 (\textit{blue}), L500 (\textit{green}) and L1000 (\textit{red}), the specifications of which are given in Table \ref{table:simulations}. Data is shown for GR (\textit{dashed lines}) and the nDGP model N1 (\textit{solid lines}).}
\label{fig:hmf_consistency}
\end{figure*}
%%%%%%%%%%%%%%%%%%%%%%%%%%%%%%%%%%%%%%%%%%%%%%%%%%

In Secs.~\ref{sec:results:concentration} and \ref{sec:results:hmf}, we combined the halo data from our DMO simulations in order to study the effects of nDGP on the halo concentration and the HMF over a wide mass range. In doing this, it is important to verify that the data from the simulations, which have different resolutions, are consistent. We therefore show, in Figs.~\ref{fig:c_consistency} and \ref{fig:hmf_consistency}, the concentration and HMF data, respectively, from each of our DMO simulations for GR and N1.

In Fig.~\ref{fig:c_consistency}, we show the binned concentration from all four of our DMO simulations, including L1000 which was excluded from our results in Sec.~\ref{sec:results:concentration}. At redshift 0, where the simulations all have sufficient resolution, the concentration follows a smooth power-law relation as a function of the mass, with the simulations showing excellent agreement at overlapping masses for both GR and N1. The agreement is not as strong at redshifts 1 and 2, where we see, for example, gaps between the L200 (blue) and L500 (green) concentrations. The concentration is slightly underestimated for haloes that are not well-resolved, affecting the data at the low-mass end (close to the lower mass cut of 2000 particles) of the L500 and L1000 data at $z=1$ and the L200 and L500 data at $z=2$. 

These resolution issues are potentially problematic for studies of the absolute concentration; however, in this work, we are more interested in the relative difference between the nDGP and GR concentration. From the lower panels of Fig.~\ref{fig:c_consistency}, it appears that the L62, L200 and L500 simulations give consistent predictions of the relative difference at overlapping masses for each redshift shown. This justifies using a halo mass cut of 2000 particles to study and model the relative difference in Sec.~\ref{sec:results:concentration}. This cut ensures that there are plenty of haloes at overlapping masses, which is important for the combined binning of the halo data, while it does not give rise to inconsistencies in the relative difference for these three simulations. We decided to exclude the L1000 simulation for a couple of reasons: the concentration suppression does not appear to be fully consistent with the data from the higher-resolution simulations -- for example, at $z=0$, the suppression in L1000 appears to be lower than the predictions from L500 at low masses and greater at high masses -- and at higher redshifts it does not have many resolved haloes.

In Fig.~\ref{fig:hmf_consistency}, we show the binned HMF from DMO simulations. The predictions of the absolute HMF, shown in the top row, agree very well. The HMF is slightly underestimated at the high-mass end of each simulation: this is a natural consequence of the limited box sizes, which causes the high-mass HMF to be incomplete. We note that combining the halo data of the four simulations and summing the total volume in the way that we have described in Sec.~\ref{sec:results:hmf} means that incompleteness is only really present for the highest-mass bins shown in Fig.~\ref{fig:hmf_combined}. The lower panels of Fig.~\ref{fig:hmf_consistency} show the relative differences between GR and N1. The predictions from the four simulations show excellent agreement, again indicating that these simulations can be safely combined.

\section{\boldmath $M_{200}$ mass function}
\label{sec:appendix:hmf}

%%%%%%%%%%%%%%%%%%%%%%%%%%%%%%%%%%%%%%%%%%%%%%%%%
\begin{figure*}
\centering
\includegraphics[width=1.0\textwidth]{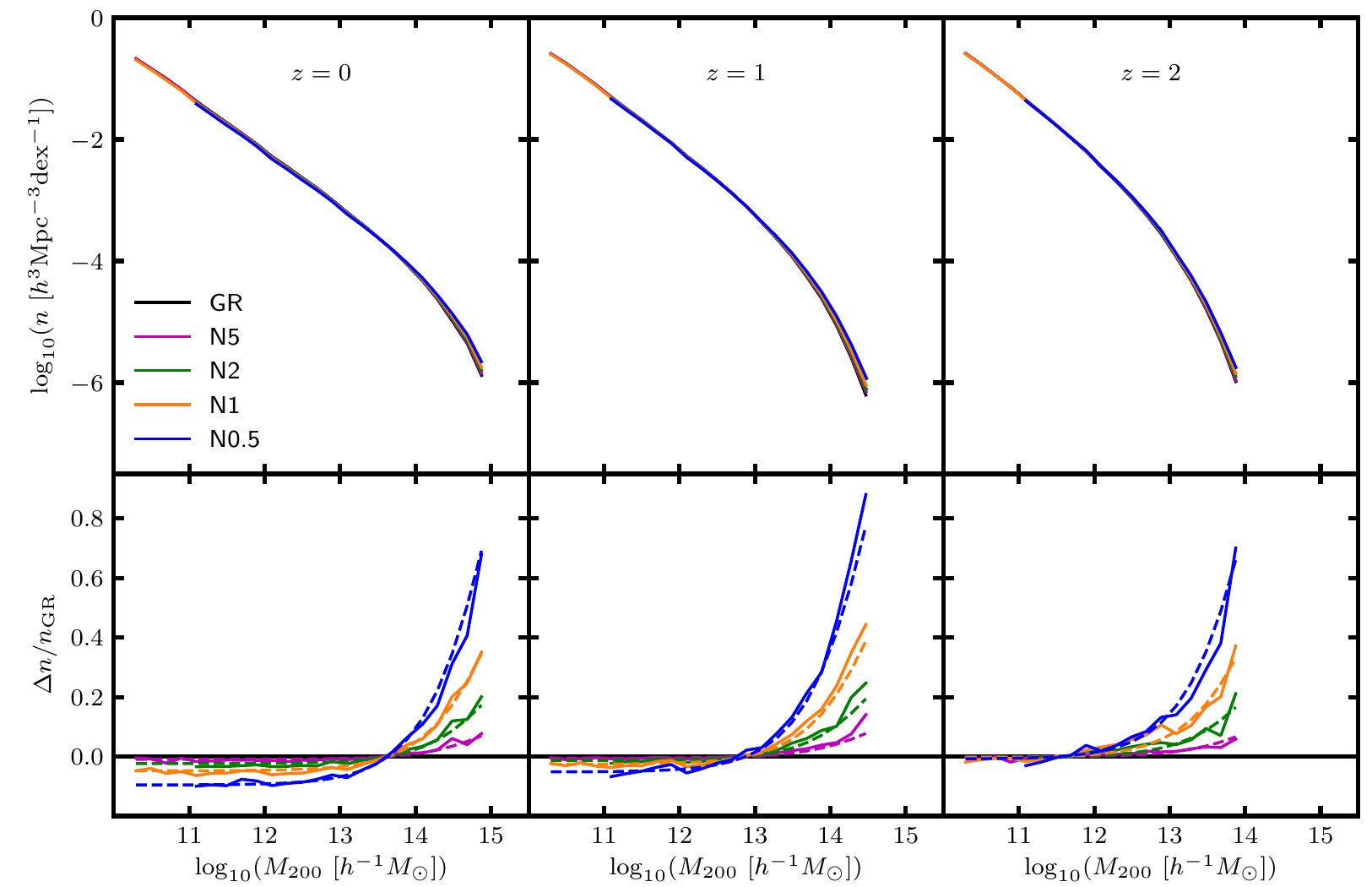}
\caption{[{\it Colour Online}] Halo mass function (\textit{top row}) and its relative difference in nDGP with respect to GR (\textit{bottom row}), as a function of the mean logarithm of the halo mass at redshifts $0$, $1$ and $2$. The results shown are similar to Fig.~\ref{fig:hmf_combined}; however, here we use mass definition $M_{200}$ instead of $M_{500}$, and the dashed lines show the predictions from the model given by Eqs.~(\ref{eq:hmf}, \ref{eq:hmf_params_200}).}
\label{fig:hmf_combined_200c}
\end{figure*}
%%%%%%%%%%%%%%%%%%%%%%%%%%%%%%%%%%%%%%%%%%%%%%%%%%

In Sec.~\ref{sec:results:hmf}, we presented our results and model for the nDGP HMF in terms of the $M_{500}$ mass definition. For completeness, we also show, in Fig.~\ref{fig:hmf_combined_200c}, the HMF in terms of the $M_{200}$ mass definition. This has again been calculated by combining the haloes from all four DMO simulations, although here we impose a lower mass threshold of 100 particles within the radius $R_{200}$ rather than $R_{500}$. We use the same set of logarithmic mass bins (with fixed width 0.2) and again show all bins that contain at least 100 haloes. 

The results in Fig.~\ref{fig:hmf_combined_200c} are very similar to Fig.~\ref{fig:hmf_combined}, with the nDGP fifth force suppressing the HMF at lower masses and enhancing the HMF at higher masses. Therefore, we are able to use the same fitting formula to model the relative difference. Replacing $M_{500}$ with $M_{200}$ in Eq.~(\ref{eq:hmf}), the best-fit parameter %functions 
are now:
\begin{equation}
    \begin{split}
    &A(H_0r_{\rm c}) = (0.59\pm0.03){\left(H_0r_{\rm c}\right)^{-1}},\\
    &B(z) = (15.22\pm0.03) - (0.441\pm0.006)z,\\
    &C(z) = (0.919\pm0.005) + (0.037\pm0.003)z.
    \end{split}
    \label{eq:hmf_params_200}
\end{equation}
The predictions of this model are also in very good agreement with the simulation measurement. As in the case of $M_{500}$, we note that this model should only be used to predict the HMF within the mass range $10^{11}h^{-1}M_{\odot}\lesssim M_{200}\lesssim M_{\rm max}(z)$, where $M_{\rm max}(z)$ is the maximum mass used for the calibration. For definition $M_{200}$, this can be estimated using:
\begin{equation}
    \log_{10}\left(M_{\rm max}M_{\odot}^{-1}h\right) = 14.93 - 0.52z.
\end{equation}
%

%%%%%%%%%%%%%%%%%%%%%%%%%%%%%%%%%%%%%%%%%%%%%%%%%%

% Don't change these lines
\bsp	% typesetting comment
\label{lastpage}
\end{document}